\documentclass[reprint,amsmath,amssymb,
            aps,pra,superscriptaddress,
            nofootinbib, floatfix,citeautoscript]{revtex4-2}

\usepackage{silence}
\usepackage[utf8]{inputenc}
\usepackage[english]{babel}
\usepackage{hyperref}
\usepackage{bm}
\usepackage{hyperref}
\usepackage{float}
\usepackage[newcommands]{ragged2e}
\usepackage{physics}
\usepackage{tabularx}
\usepackage{mathtools}
\usepackage[toc,page,header]{appendix}
\usepackage{nicefrac}
\usepackage{sidecap}
\usepackage{multibib}
\usepackage[dvipsnames]{xcolor}
\usepackage{booktabs}
\usepackage{float}
\usepackage{graphicx}
\usepackage[capitalise]{cleveref}
\usepackage[version=4]{mhchem}
\usepackage{tabularx}
\usepackage{ulem}
\usepackage{enumitem}
\newcolumntype{C}[1]{>{\centering\let\newline\\\hspace{0pt}}m{#1}}

\usepackage[per-mode=power, 
            exponent-product = \times, 
            inter-unit-product = \ensuremath{{\cdot}}, 
            tight-spacing = true,
            number-unit-product=~,
            parse-numbers=false,]{siunitx}
\AtBeginDocument{\RenewCommandCopy\qty\SI}
\ExplSyntaxOn
\msg_redirect_name:nnn { siunitx } { physics-pkg } { none }
\ExplSyntaxOff

\usepackage[
    protrusion=true,
    activate={true,nocompatibility},
    final,
    tracking=true,
    kerning=true,
    spacing=true,
    factor=1000]{microtype}
\hypersetup{colorlinks=true,  linkcolor=blue, 
            citecolor=blue, filecolor=blue,  
            urlcolor=blue}
\usepackage[raggedright]{titlesec}
\usepackage{caption}
\titleformat*{\section}{\bfseries\large}
\titleformat*{\subsection}{\bfseries}
\titlespacing{\section}{0em}{1em}{0.25em}
\renewcommand{\textcite}[1]{\citenum{#1}}
\Crefformat{appendix}{Supplementary Information #2#1#3}

\DeclareSIUnit{\rad}{rad}
\begin{document}

\author{Gr\'egory Moille}%
\email{gmoille@umd.edu}
\affiliation{Joint Quantum Institute, NIST/University of Maryland, College Park, USA}
\affiliation{Microsystems and Nanotechnology Division, National Institute of Standards and Technology, Gaithersburg, USA}
\author{Sashank~Kaushik~Sridhar}
\affiliation{Department of Mechanical Engineering, University of Maryland, College Park, USA}
\author{Pradyoth~Shandilya}
\affiliation{University of Maryland, Baltimore County, Baltimore, MD, USA}
\author{Avik~Dutt}
\affiliation{Department of Mechanical Engineering, University of Maryland, College Park, USA}
\affiliation{Institute for Physical Science and Technology, University of Maryland, College Park, USA}
\affiliation{National Quantum Laboratory (QLab) at Maryland, College Park, USA}
\author{Curtis~Menyuk}
\affiliation{University of Maryland, Baltimore County, Baltimore, MD, USA}
\author{Kartik~Srinivasan}
\affiliation{Joint Quantum Institute, NIST/University of Maryland, College Park, USA}
\affiliation{Microsystems and Nanotechnology Division, National Institute of Standards and Technology,
Gaithersburg, USA}
\date{\today} 

\newcommand{\mytitle}{Toward Chaotic Group Velocity Hopping of an On-Chip Dissipative Kerr Soliton}
 \title{\mytitle}

 \begin{abstract}
        Chaos enables randomness-based applications, particularly in photonic systems. Integrated optical frequency combs (microcombs) have previously been observed in either chaotic modulation instability or stable, low-noise dissipative Kerr soliton (DKS) regimes. 
        In this work, we demonstrate a new microcomb state where a single DKS exhibits chaotic behavior. %
        By phase modulating the Kerr-induced synchronization (KIS) between a DKS and an externally injected reference laser, we observe chaotic group velocity hopping of the soliton, causing random transitions of the repetition rate. Using a chip-integrated octave-spanning microcomb, we experimentally validate the second-order Adler equation describing KIS, allowing us to predict and demonstrate this chaotic DKS hopping. This work connects nonlinear dynamics with optical soliton physics, providing a deterministic framework for triggering microcomb chaos in the solitonic state.
    \end{abstract}
 
\maketitle



\section{Introduction}
\begin{figure}
     \centering{
     \includegraphics[width = \columnwidth]{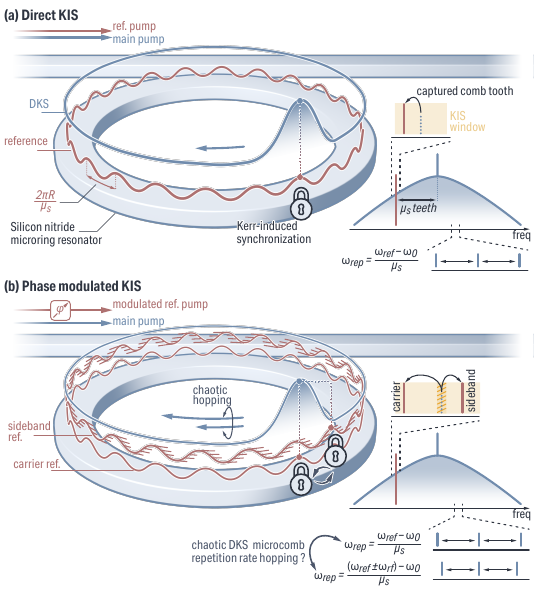}}
     \caption{\label{fig:1}
\textbf{Kerr-induced synchronization (KIS): Direct and phase-modulated versions --- }%
\textbf{a}~Direct KIS occurs when a weak continuous-wave reference laser enters a microring resonator containing a dissipative Kerr soliton (DKS) pumped by a main laser. %
The Kerr nonlinearity synchronizes the reference and DKS when their phase difference is small, dual-pinning the microcomb repetition rate $\omega_\mathrm{rep}$ through the capture of one comb tooth. %
\textbf{b}~Under phase modulation, the DKS can lock to multiple reference attractors with distinct walk-off, hence leading to different group velocity locking, correspondingly a comb tooth locks to either the carrier or a sideband, causing an abrupt $\omega_\mathrm{rep}$ change. %
We examine the Adler model accuracy for KIS, predicting chaotic DKS $\omega_\mathrm{rep}$ hopping.
}
 \end{figure}

Chaos manifests throughout nature: in fluid turbulence~\cite{SreenivasanRev.Mod.Phys.1999}, biological rhythms~\cite{Heltbergcels2021}, electronic circuits~\cite{MatsumotoIEEETrans.CircuitsSyst.1984}, and famously, atmospheric convection~\cite{LorenzJ.AtmosphericSci.1963}. Mathematically, chaos occur in autonomous systems of three nonlinearly coupled independent variables, where small differences in initial conditions---inherent to any real world system---leads to diverging outcomes, making future prediction impossible. Photonic chaos has been studied in lasers with feedback~\cite{RyanIEEEJ.QuantumElectron.1994} and mode-locked systems~\cite{AkhmedievPhys.Rev.E2001}. Rather than an obstacle, optical chaos enables high-speed random number generation~\cite{ReidlerPhys.Rev.Lett.2009}, secure communication~\cite{ArgyrisNature2005}, and reservoir computing~\cite{RafayelyanPhys.Rev.X2020}.\\
\indent In integrated frequency combs, chaotic dynamics through modulation instability (MI)---arising from a continuous wave (CW) laser pumping an on-chip Kerr microresonator---has applications in ranging~\cite{LukashchukAPLPhotonics2023}, optical decision making~\cite{LukashchukAPLPhotonics2023}, and distributed encrypted communication~\cite{MorenoOpt.ExpressOE2024}. %
Yet, most applications also benefit from the creation of a fixed grid of frequency markers, typically obtained from the creation of dissipative Kerr solitons (DKSs)~\cite{KippenbergScience2018}. %
As unique solutions of the Lugiato-Lefever equation (LLE)~\cite{ChemboPhys.Rev.A2013}, DKSs exhibit robustness against perturbation and serve as the cornerstone for deployable metrology systems~\cite{DiddamsScience2020a}. %
Therefore, to simultaneously leverage equidistant frequency markers and chaotic behavior, optical cavity solitons require an additional coupling mechanism. %
This manifests in soliton bunches or molecules, where internal dynamics can drive chaotic interactions~\cite{GreluNaturePhoton2012,LecaplainPhys.Rev.Lett.2012,SongLaserPhotonicsRev.2023}. %
Yet, such behavior remains elusive in integrated photonics, especially in the single DKS regime driven by a single pump.\\
\indent To access chaos in a single on-chip DKS, we phase-lock the soliton to an externally injected reference laser using Kerr-induced synchronization (KIS). %
In KIS the reference captures one comb tooth [\cref{fig:1}a], doubly pinning the microcomb~\cite{MoilleNature2023,WildiAPLPhotonics2023}. %
This creates a unique DKS attractor that bypasses intrinsic microresonator fluctuations~\cite{Moille2024_arXivTRN}. %
KIS differs significantly from other soliton synchronization mechanisms~\cite{JangNatCommun2015a,JangNaturePhoton2018b,EnglebertNat.Phys.2023,ColePhys.Rev.Lett.2019}. %
While these other mechanisms can be described by a first-order Adler equation---corresponding only to the overdamped case---KIS can only be described by an extended second-order Adler equation. %
Importantly, this second-order equation describes a physical coupling mechanism through phase modulation that enables a path to chaos.
However, thus far this model has only demonstrated qualitative agreement in the KIS DC regime~\cite{YangNaturePhoton2017, MoilleNature2023,Sun2024}, and its second-order temporal derivative---crucial for chaotic DKS---remains experimentally unverified.\\
\indent Here, we demonstrate the second-order Adler equation's validity for KIS using an integrated octave-spanning microcomb. %
We quantitatively validate the model against experiments and Lugiato-Lefever simulations in reference carrier and sideband synchronization regimes, while also demonstrating sub-harmonic locking---a phenomenon unique to a high-order Adler equation~\cite{ValizadehJNMP2008}. %
Based on these insights, we leverage the Adler model's predictions to experimentally show chaotic DKS hopping between reference and sideband strange attractors, resulting in chaotic group velocity hopping that appears as random jumps of the microcomb repetition rate [\cref{fig:1}b].

 \section{Phase-Modulated Adler Model}
 
 Our system consists of a microring resonator driven by a CW main pump that generates a DKS. %
 The DKS circulates through the resonator, with a portion periodically extracted into the bus waveguide, creating a temporal pulse train at rate $\omega_\mathrm{rep}$, corresponding to a frequency comb in the spectral domain. %
 A weak CW reference pump, injected into the resonator at a frequency close to a DKS comb tooth, interacts with the DKS through the $\chi^{(3)}$ nonlinearity. %
 This interaction phase-locks the DKS, synchronizing both group and phase velocities of the DKS and reference intracavity field. %
 Equivalently, the reference pump captures the nearest DKS comb tooth, while the dual-pinning by the main and reference pump frequencies disciplines the comb's repetition rate [\cref{fig:1}a]. %
 This synchronization mechanism obeys a second-order Adler equation that can be derived from the multi-pump Lugiato-Lefever equation~\cite{TaheriEur.Phys.J.D2017} (see Supplementary Material \textbf{[URL]} which includes~\cite{CoulletAmericanJournalofPhysics2005,MoilleOpt.Lett.2019,ZhangOptica2019,ZhouLightSciAppl2019,StonePhys.Rev.Lett.2020,WangOptica2017a,Rukhin2010,RandomGithub}, and ref~\cite{MoilleNature2023}):
 \begin{equation}
     \label{eq:dc_adler}
        \beta\frac{\partial^2 \varphi(\tau)}{\partial \tau^2} + \frac{\partial \varphi(\tau)}{\partial \tau} =  \alpha + \sin\left( \varphi \right).
 \end{equation}

 \noindent with the normalized time $\tau=\varOmega_0 t$, where $\varOmega_0=2\mu_s D_1 E_\mathrm{kis}$ is the half-synchronization KIS window. %
 Here, $D_1$ is the free spectral range around the main pump, and $\mu_s$ is the modal separation between main and reference pump. %
 The KIS energy $E_\mathrm{kis} = \sqrt{E_0(\mu_s) P_\mathrm{ref}\kappa_\mathrm{ext}}/\kappa E_\mathrm{dks}$ depends on the soliton energy $E_\mathrm{dks}$, comb tooth energy $E_0(\mu_s)$ at synchronization mode $\mu_s$, reference pump power $P_\mathrm{ref}$, and total and coupling loss rates $\kappa$ and $\kappa_\mathrm{ext}$, respectively. %
 The phase difference $\varphi$ (modulo $2\pi$) is between the DKS and reference color. %
 $\varOmega = \Delta\omega_\mathrm{ceo}/\varOmega_0$ represents the normalized frequency offset between reference and closest comb tooth, vanishing at synchronization. %
 The synchronization detuning factor is $\alpha = \varpi/\varOmega_0$, with $\varpi$ being the reference frequency offset from natural comb tooth frequency. %
 The McCumber coefficient~\cite{McCumberJ.Appl.Phys.1968} $\beta = \varOmega_0/\kappa$ characterizes system bistable dynamics.
  
In the phase-modulated reference regime [\cref{fig:1}b], \cref{eq:dc_adler} becomes (Supplementary Material S.2):

 \begin{equation}
     \label{eq:ac_adler}
     \left \{
     \begin{aligned}
        \frac{\partial\psi(\tau)}{\partial \tau}  &= \varOmega_\mathrm{ext},\\ 
        \frac{\partial\varphi(\tau)}{\partial \tau}&= \varOmega,\\
         \beta\frac{\partial \varOmega(\tau)}{\partial \tau} &= - \varOmega  + \alpha + \sin(\varphi+ A\cos\left( \psi \right)).
     \end{aligned}
     \right . 
 \end{equation}
 
 Here, $\varOmega_\mathrm{ext} = \omega_\mathrm{rf}/\varOmega_0$ denotes the normalized phase modulation frequency. %
 \cref{eq:ac_adler} now exhibits a fixed point at $\alpha \approx n \varOmega$ for any $\beta$, generating (not necessarily circular) closed orbits in the $\{\varphi, \varOmega\}$ space. %
 Hence, through the introduction of the reference phase modulation, \cref{eq:ac_adler} becomes an autonomous system of three nonlinearly coupled independent variables, and can exhibit, among other nonlinear strange attractors~\cite{PhilipPhil.Trans.R.Soc.Lond.A1979}, chaotic behavior.

 \section{Model Verification}
 
 \begin{figure}[t]
     \begin{center}
         \includegraphics[width = \columnwidth]{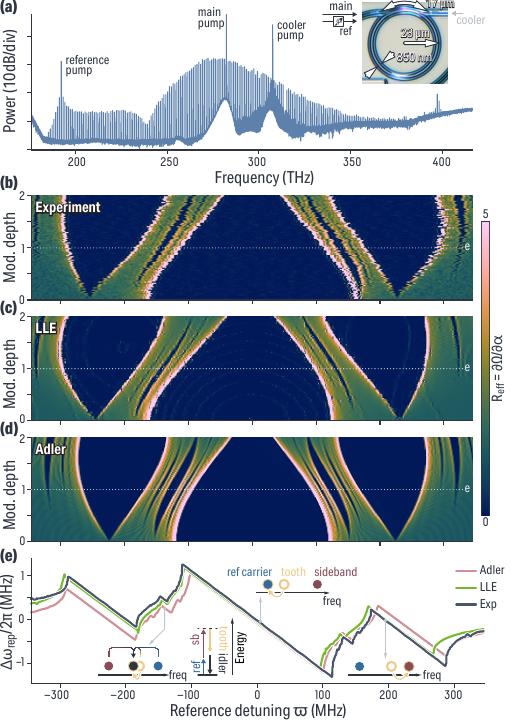}    
     \end{center}
     \caption{\label{fig:2}
\textbf{Adler model verification from repetition rate entrainment --- }%
\textbf{a} Octave-spanning microcomb from the integrated microresonator (inset), pumped at \qty{\omega_0/2\pi=282.3}{\THz} with the KIS reference at $\mu_s = -90$ (\qty{\omega_\mathrm{ref}/2\pi= 192.5}{\THz}). %
  A cooler laser at \qty{307.6}{\THz} thermally stabilizes the resonator. %
\textbf{b} Mapping of $R_\mathrm{eff} = \partial \varOmega / \partial\varpi$ with reference detuning $\varpi$ and modulation depth $A$, highlighting DKS comb tooth capture when $R_\mathrm{eff}$ vanishes. %
\textbf{c} LLE modeling, and  
\textbf{d} Adler equation modeling, showing the reduced model accurately captures the system dynamics.  
\textbf{e} Repetition rate disciplining from experiment (blue), LLE (green), and Adler (pink) at $A=1$ (dotted line in b-c) shows excellent agreement. %
The disciplining occurs at the reference, its sideband, and phase-modulation subharmonics via four-wave mixing, as in the schematic.
     }
 \end{figure}

\begin{figure}[!t]  
    \centering
    \includegraphics[width = \columnwidth]{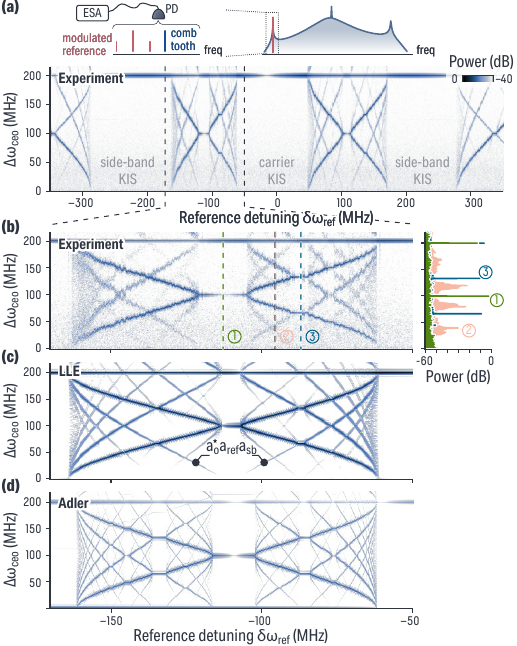}
    \caption{\label{fig:3}
\textbf{Adler model verification from the dynamical spectrogram --- } %
\textbf{a}~Homodyne $\Delta\omega_\mathrm{ceo}$ spectrogram obtained for (\qty{\omega_\mathrm{rf}/2\pi=200}{\MHz}and $A=1.3$).  %
\textbf{b}~Left: zoom-in between detuning $-\omega_\mathrm{rf}<\varpi<0$, highlighting subharmonic locking. %
Right: fixed-detuning spectra, showing the low-noise regime of sub-harmonic locking mediated by four-wave mixing (\raisebox{.5pt}{\textcircled{\raisebox{-.9pt}{1}}} and \raisebox{.5pt}{\textcircled{\raisebox{-.9pt}{3}}}) while the unsynchronized state (\raisebox{.5pt}{\textcircled{\raisebox{-.9pt}{2}}}) yield broader linewidths from optical beating between the comb tooth and the reference. %
\textbf{c} LLE and \textbf{d} Adler equation simulations eriments, confirming the Adler model's validity.
    }
\end{figure}

On its own, \cref{eq:dc_adler} only allows for qualitative verification of the KIS frequency width $\varOmega_0$, and synchronization does not necessarily require a second-order Adler equation. %
By contrast, \cref{eq:ac_adler} supports far richer dynamics. %
Equation~(\ref{eq:ac_adler}) is equivalent to the resistively and capacitively shunted junction (RCSJ) model of an AC-driven Josephson junction~\cite{JosephsonPhysicsLetters1962a,ValizadehJNMP2008}, where sub-harmonic synchronization emerges from the second-order derivative~\cite{ValizadehJNMP2008}. %
The experimental observation of such behavior would validate the necessity of a second-order Adler equation and suggest that its other predictions---such as chaotic behavior---may be experimentally accessible.\\
\indent Our system, detailed in Supplementary Material S.3, consists of a \qty{23}{\um} radius \ce{Si3N4} microring resonator generating an octave-spanning DKS microcomb~[\cref{fig:2}a]. %
It is synchronized with a $\qty{\omega_\mathrm{rf}/2\pi = 200}{\MHz}$ modulated reference pump at azimuthal mode $\mu_s = -90$ ($\qty{192.5}{\THz}$). %
We measure the microcomb instantaneous repetition rate change $\Delta \omega_\mathrm{rep}$ against the natural reference detuning $\varpi$ (i.e., the detuning relative to the unsynchronized comb tooth). %
KIS occurs for the same optical frequency division (OFD) factor $\mu_s$  as in the unmodulated KIS case for $R_\mathrm{eff} = \mu_s\partial \Delta\omega_\mathrm{rep}/\partial \varpi - 1\equiv \partial\varOmega/\partial\alpha=0$. %
As expected, synchronization occurs around the reference pump and its sideband [\cref{fig:2}b]. %
The size of the KIS window scales with the absolute values of the first order Bessel functions $J_0(A)$ and $J_1(A)$, corresponding to the amplitude of the reference carrier and sideband from the expansion of \cref{eq:ac_adler}. %
We also observe the sub-harmonic synchronization tongues, known as Shapiro-steps in superconducting systems~\cite{ShapiroPhys.Rev.Lett.1963}, emerging at $\varpi \approx p\omega_\mathrm{ext}/m$, with observations up to $m=3$, which are well-known in AC driven Josephson junctions.\\
 \indent We numerically reproduce the experimental result [\cref{fig:2}c] by extracting the low-pass filtered DKS repetition rate using the modified Lugiato-Lefever equation~\cite{TaheriEur.Phys.J.D2017,Menyuk2024,TaheriLaserPhotonicsRev.2025} that accounts for the reference modulation (Supplementary Material S.1).
 \begin{equation}
     \label{eq:LLE}
     \begin{split}
     \frac{\partial a(\theta,t)}{\partial t} &= -\left(\frac{\kappa}{2} - i\delta\omega_0\right)a + i\mathcal{D}(a) 
         -i\gamma|a|^2 a
         + iF_{0}\\
     &+ iF_\mathrm{ref}\mathrm{exp}\Big\{ i\big[\varpi t + A\cos\left(\omega_\mathrm{rf} t\right) \big] + i\mu_s \theta \Big\}
     \end{split}
 \end{equation}

\noindent  with the resonator parameters (Supplementary Material S.1) which are the effective nonlinearity $\gamma$ , the total and coupling loss rates $\kappa$ and $\kappa_\mathrm{ext}$, and the dispersion operator $\mathcal{D}(a)$. %
$A(\mu,t)$ is the azimuthal Fourier transform of $a(\theta,t)$, %
$\delta\omega_0$ is the main pump detuning, and $F_0=\sqrt{\kappa_\mathrm{ext}P_0}$ and $F_\mathrm{ref}=\sqrt{\kappa_\mathrm{ext}P_\mathrm{ref}}$ represent the main pump and reference energy based on experimental power measurements. %
In the first sideband approximation, the reference modulation expands as the sum of two driving fields $iJ_0(A)F_\mathrm{ref}\mathrm{exp}\left( {i\varpi t + i\mu_s \theta} \right)$ and $iJ_1(A)F_\mathrm{ref}\mathrm{exp}\left( {i(\varpi \pm \omega_\mathrm{rf})t + i\mu_s \theta} \right)$. %
Outside synchronization from the carrier ($\varpi\approx 0$) or sideband ($\varpi \pm \omega_\mathrm{rf} \approx 0$), the system forms a multi-color soliton~\cite{Menyuk2024} with components $a_0$ from the DKS, $a_\mathrm{ref}$ from the reference carrier and $a_\mathrm{sb}$ from the sideband. %
These components travel at identical angular group velocity but different angular phase velocities. %
The system can be decomposed through a multi-color expansion $a = a_0 + a_\mathrm{ref}\mathrm{e}^{i\mu_s \theta - i\varpi t} + a_\mathrm{sb} \mathrm{e}^{i\mu_s \theta - i(\varpi + \omega\mathrm{rf}) t}$~\cite{Menyuk2024,MoilleNat.Photon.2025}, which after reinjection in~\cref{eq:LLE} yields the master equation for the soliton (Supplementary Material S.1):
 \begin{equation}
     \begin{split}
         \frac{\partial a_0}{\partial t} =& -\left(\frac{\kappa}{2} - i\delta\omega_0\right)a_0 + i\mathcal{D}(a_0) \\
         &-i\gamma(|a_0|^2 + 2|a_\mathrm{ref}|^2 + 2|a_\mathrm{sb}|^2) a_0 \\
         &- 2i\gamma a_0^*a_\mathrm{ref}a_\mathrm{sb} \mathrm{e}^{i(2\varpi \pm \omega_\mathrm{rf})t + i\mu_s\theta}
     + iF_{0}
     \end{split}
 \end{equation}
 
 \noindent The parametric tone $2\gamma a_0^*a_\mathrm{ref}a_\mathrm{sb}$, produced by four-wave mixing between the DKS, reference, and sideband, generates an idler that synchronizes the DKS when their phase difference approaches zero ($2\varpi \pm \omega_\mathrm{rf} \approx 0$). %
 This synchronization occurs at reference detuning $\varpi \approx \pm\omega_\mathrm{rf}/2$, analogous to the recently observed parametric-KIS~\cite{MoilleArXivParametric2024}, in which synchronization occurs without direct comb tooth capture by an injected reference. %
 The model extends to higher sideband orders, explaining sub-harmonic synchronization at $\varpi \approx \pm n\omega_\mathrm{rf}/p$ in simulations and experiments. %
 Finally, we verify the Adler model in~\cref{eq:ac_adler}, normalizing it from the experimental value of the KIS half-window \qty{\varOmega_0 = 130}{\MHz} at $A=0$ and \qty{\kappa=200}{\MHz} (\textit{i.e $\beta = 0.65$}) which, along with $\omega_\mathrm{rf}$, is sufficient to define all parameters of~\cref{eq:ac_adler}. We compare these results against both the LLE and experiment [\cref{fig:2}d] by extracting the low-passed filter $\Delta\omega_\mathrm{rep}$ at $A=1$, showing excellent agreement. The repetition rate disciplining occurs at the reference, its sideband, and subharmonics of the phase modulation frequency from nonlinear interaction. The experimental repetition rate disciplining is accurately captured by both the LLE and Adler models [\cref{fig:2}e].\\
\indent  Low-pass filtered detection of $\Delta\omega_\mathrm{rep}$, from which we derive $R_\mathrm{eff}$, helps us visualize the fixed point and subharmonic synchronization against modulation depth $A$. %
However, it does not fully confirm that the KIS dynamics follow the phase modulated Adler model. %
Because \cref{eq:ac_adler} represents a driven system, its stable points appear as fixed points under $2\pi/\omega_\mathrm{rf}$ stroboscopic observations (or via low-pass filtering as previously described) but, for any other observation, manifest as a closed orbit or a limit cycle for synchronized and unsynchronized cases, respectively [\cref{fig:3}a]. %
Hence, Fourier analysis of $\varOmega$ will allow us to verify the Adler model. %
Experimentally, we capture the full dynamics within the \qty{\omega_\mathrm{rf}/2\pi=200}{\MHz} bandwidth by directly measuring the beat frequency between the reference laser and its nearest comb tooth for a modulation depth $A=1.33$. %
 %
From the spectrogram of $\Delta\omega_\mathrm{ceo} = \varOmega \times \varOmega_0$, the two synchronization regimes appear: direct synchronization and sub-harmonic locking. Direct synchronization from the carrier or sideband produces a single tone at the spinning frequency $\omega_\mathrm{rf}$ of the closed orbit, while sub-harmonic locking generates harmonics at $p\times\omega_\mathrm{rf}/n$ (notably at $p\times\omega_\mathrm{rf}/3$). This latter state shows reduced noise compared to the unsynchronized regime~[\cref{fig:3}b], indicating mode-locking despite the absence of direct comb tooth capture. Both the LLE modeling and the Adler model of \cref{eq:ac_adler} reproduce the complete spectrogram, with the latter also capturing all frequencies in the limit cycle case outside synchronization.

 \section{Chaotic group velocity hopping}
 \begin{figure}[!t]  
    \centering
    \includegraphics[width = \columnwidth]{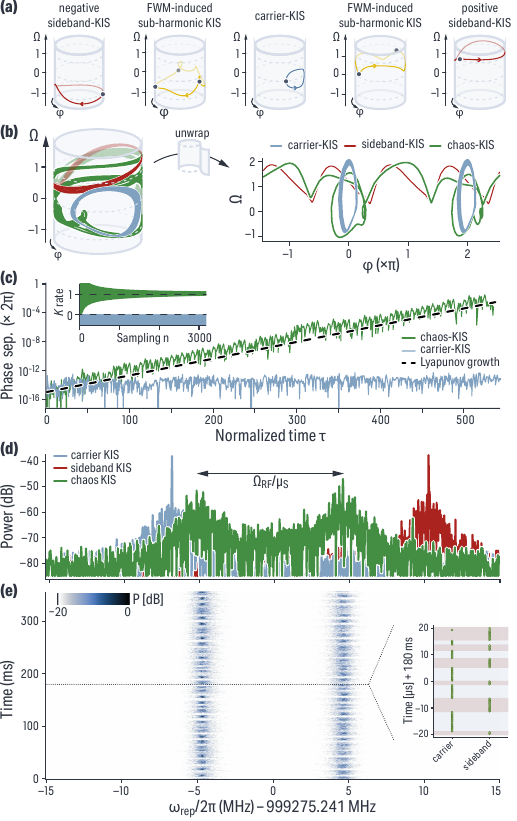}
\caption{\label{fig:4}
\textbf{Toward stable-DKS chaotic behavior --- }%
\textbf{a}~Adler simulation of the modulated reference system highlights attractors in $\{\varphi, \varOmega\}$ space. As $\varphi$ is $2\pi$-periodic, the projection is cylindrical. %
Solid lines show closed orbits at $\varOmega_\mathrm{ext}$; markers show $2\pi/\varOmega_\mathrm{ext}$ stroboscopic observation highlighting synchronizations. %
\textbf{b} Phase modulated Adler simulation for $\alpha = 0.58$ shows two coexisting attractors (green) orbiting around $\varOmega=0$ (blue, carrier KIS) and $\varOmega=1$ (red sideband KIS) leading to the chaotic behavior.  %
\textbf{c}~Phase separation between initially offset systems ($\Delta\varphi=10^{-15}$) reveals chaos via exponential growth and maximal Lyapunov exponent $\lambda=2\pi \times 5.33\times 10^{-2}$. %
The $0\text{-}1$ test confirms chaos ($K=1$), unlike direct-carrier-KIS (no growth and $K=0$). %
 \textbf{d}~Experimental RF spectrum for carrier-KIS (blue) and sideband-KIS (red). In chaos, dual repetition rate DKS appears, separated by the OFDed modulation \qty{\omega_\mathrm{rf}/\mu_s=2\pi\times9.44}{\MHz}. %
 \textbf{e}~The spectrogram confirms stable-DKS chaos, showing hopping between the two repetition rates. %
 Inset: zoom at \qty{180}{\ms} reveals random synchronization states.
}
\end{figure}

Having shown that the second-order Adler equation accurately describes the soliton phase dynamics under KIS, we now apply this model to predict the regime in which chaotic behavior of the DKS is expected. %
Geometrically, it is intuitive to view both chaos and the regimes studied in the preceding sections in the ($\varphi,\varOmega$) space~[\cref{fig:4}a], cylindrical since $\varphi$ is $2\pi$ periodic while $\varOmega$ is unbounded~\cite{Strogatz2019}. %
Computational solution of~\cref{eq:ac_adler} confirms that for $\alpha = 0.58 $, the strange attractor orbiting orbiting around $\varOmega=0$ with a null winding number in the $\varphi$ revolution co-exists with the strange attractor with unitary winding number orbiting around $\varOmega=1$~[\cref{fig:4}b], and in contrast to the cases in \cref{fig:4}a that display only a single attractor. %
We confirm the chaotic behavior by simulating two systems with identical parameters with the exception of an offset in initial condition by $\Delta\varphi=10^{-15}$, from which we can extract a phase separation at all normalized times $\tau$~[\cref{fig:4}c]. %
We extract the maximal Lyapunov exponent~\cite{WolfPhysicaD:NonlinearPhenomena1985}, yielding a growth rate of \qty{2\pi\times5.33\times10^{-2}}{}, hence greater than zero and confirming chaotic behavior, in contrast to standard KIS (yielding a zero growth). %
We also perform a $0-1$ test~\cite{GottwaldSIAMJ.Appl.Dyn.Syst.2009} that confirms the chaotic behavior under the chosen conditions. \\
\indent We demonstrate chaotic KIS behavior experimentally by setting \qty{\omega_\mathrm{rf}/2\pi=850}{\MHz}, which is about the half-KIS window $\varOmega_0$ for our system and a  reference pump power \qty{P_\mathrm{ref}=300}{\uW}. %
At the boundary between carrier and sideband-KIS regimes, chaos emerges as both strange attractors coexist. %
Since each attractor corresponds to a different comb tooth capture process (the carrier or the sideband), each attractor correspond to a fixed microcomb repetition rate. %
From simulations, this chaotic region spans nearly \qty{100}{\MHz}, and is ample for easy laser frequency tuning and parking.  %
The chaotic transition between each attractor creates a repetition rate jump, appearing in the spectrum analyzer as coexisting tones separated by $\omega_\mathrm{rf}/\mu_s$ [\cref{fig:4}d]. %
A spectrogram recording shows the temporal transitions between repetition rates [\cref{fig:4}e]. %
From this data, we perform a chi-square test (see supplementary S.4) yielding p-values of $p=0.81$ and $p=0.96$ for $-1$ and $+1$ random walk, and confirming the chaotic behavior under such specific KIS conditions.

\section{Discussion}
In conclusion, our combined experimental and theoretical work quantitatively verifies that the second-order Adler model accurately describes KIS DKS dynamics. %
Moving from DC behavior to a phase-modulated regime, we observe unique subharmonic synchronization from the second-order KIS Adler model. This verification confirms the Adler model as a predictive tool for DKS behavior. %
Specifically, the phase-modulated Adler model predicts conditions for chaotic behavior intuitively and with much less computational cost than full LLE modeling. %
We experimentally confirm this by chaotic repetition rate hopping of an octave-spanning DKS microcomb at a phase modulation comparable to the KIS half-window. %
Our work presents a new tool for generating optical chaos on a low-power, on-chip platform, harnessing a single low-noise DKS with the secure randomness of chaotic systems. %
By showing that chaotic behavior in microcombs extends beyond MI states, we enable further studies of novel DKS regimes. %
While our demonstration uses a phase-modulated injected reference, the Adler model remains agnostic to which oscillator is modulated, suggesting these observations would apply to breather states. %
Beyond chaos, our verification that KIS follows a second-order Adler equation provides insight into complex behaviors, including indirect comb tooth capture via coherent four-wave mixing, and may clarify unexplained frequency comb or soliton synchronization observations~\cite{ShortissOpt.Express2019,Wu2025}. %
The second-order Adler equation also describes short Josephson junctions~\cite{ValizadehJNMP2008}, suggesting potential cross-field insights similar to developments in on-chip optical synthetic dimensions~\cite{YuanOptica2018,BellOptica2017,BalcytisSci.Adv.2022,SridharNat.Phys.2024a} and optical topological systems~\cite{LinNatCommun2016a,LustigNature2019,LustigAdv.Opt.Photon.AOP2021,EnglebertNat.Phys.2023}. %
This approach may overcome synthetic frequency lattice limitations imposed by the fast round-trip time of integrated cavity solitons. %

 \section{Acknowledgments}
  K.S. and G.M acknowledge 
 support from the Space Vehicles Directorate of the Air Force Research Laboratory and the NIST-on-a-chip program. P.S and C.M. acknowledges support from the National Science Foundation (grant ECCS-1807272), Air Force Office of Scientific Research grant (Grant FA9550-20-1-0357), and a collaborative agreement with the National Center for Manufacturing Sciences (2022138-142232) as a sub-award from the US Department of Defense (cooperative agreement HQ0034-20-2-0007). The Scientific colour map batlow~\cite{Crameri2023} is used in this study to prevent visual distortion of the data and exclusion of readers with colour-vision deficiencies~\cite{CrameriNatCommun2020}.
The authors thanks Sanzida Akter and Shao-Chien Ou for helpful discussions. %
 G.M thanks T.B.M. %
 \noindent The data supporting the plots and findings in this paper are available from the corresponding authors upon reasonable request. The code, modified from a public source~\cite{MoilleJ.RES.NATL.INST.STAN.2019}, can be obtained from the corresponding author. %
 G.M., C.M. and K.S have submitted a provisional patent application based on aspects of the work presented in this paper. %
 Correspondence and requests for materials should be addressed to G.M.


\begin{thebibliography}{59}%
\makeatletter
\providecommand \@ifxundefined [1]{%
 \@ifx{#1\undefined}
}%
\providecommand \@ifnum [1]{%
 \ifnum #1\expandafter \@firstoftwo
 \else \expandafter \@secondoftwo
 \fi
}%
\providecommand \@ifx [1]{%
 \ifx #1\expandafter \@firstoftwo
 \else \expandafter \@secondoftwo
 \fi
}%
\providecommand \natexlab [1]{#1}%
\providecommand \enquote  [1]{``#1''}%
\providecommand \bibnamefont  [1]{#1}%
\providecommand \bibfnamefont [1]{#1}%
\providecommand \citenamefont [1]{#1}%
\providecommand \href@noop [0]{\@secondoftwo}%
\providecommand \href [0]{\begingroup \@sanitize@url \@href}%
\providecommand \@href[1]{\@@startlink{#1}\@@href}%
\providecommand \@@href[1]{\endgroup#1\@@endlink}%
\providecommand \@sanitize@url [0]{\catcode `\\12\catcode `\$12\catcode `\&12\catcode `\#12\catcode `\^12\catcode `\_12\catcode `\%12\relax}%
\providecommand \@@startlink[1]{}%
\providecommand \@@endlink[0]{}%
\providecommand \url  [0]{\begingroup\@sanitize@url \@url }%
\providecommand \@url [1]{\endgroup\@href {#1}{\urlprefix }}%
\providecommand \urlprefix  [0]{URL }%
\providecommand \Eprint [0]{\href }%
\providecommand \doibase [0]{https://doi.org/}%
\providecommand \selectlanguage [0]{\@gobble}%
\providecommand \bibinfo  [0]{\@secondoftwo}%
\providecommand \bibfield  [0]{\@secondoftwo}%
\providecommand \translation [1]{[#1]}%
\providecommand \BibitemOpen [0]{}%
\providecommand \bibitemStop [0]{}%
\providecommand \bibitemNoStop [0]{.\EOS\space}%
\providecommand \EOS [0]{\spacefactor3000\relax}%
\providecommand \BibitemShut  [1]{\csname bibitem#1\endcsname}%
\let\auto@bib@innerbib\@empty
\bibitem [{\citenamefont {Sreenivasan}(1999)}]{SreenivasanRev.Mod.Phys.1999}%
  \BibitemOpen
  \bibfield  {author} {\bibinfo {author} {\bibfnamefont {K.~R.}\ \bibnamefont {Sreenivasan}},\ }\bibfield  {title} {\bibinfo {title} {Fluid turbulence},\ }\href {https://doi.org/10.1103/RevModPhys.71.S383} {\bibfield  {journal} {\bibinfo  {journal} {Reviews of Modern Physics}\ }\textbf {\bibinfo {volume} {71}},\ \bibinfo {pages} {S383} (\bibinfo {year} {1999})}\BibitemShut {NoStop}%
\bibitem [{\citenamefont {Heltberg}\ \emph {et~al.}(2021)\citenamefont {Heltberg}, \citenamefont {Krishna}, \citenamefont {Kadanoff},\ and\ \citenamefont {Jensen}}]{Heltbergcels2021}%
  \BibitemOpen
  \bibfield  {author} {\bibinfo {author} {\bibfnamefont {M.~L.}\ \bibnamefont {Heltberg}}, \bibinfo {author} {\bibfnamefont {S.}~\bibnamefont {Krishna}}, \bibinfo {author} {\bibfnamefont {L.~P.}\ \bibnamefont {Kadanoff}},\ and\ \bibinfo {author} {\bibfnamefont {M.~H.}\ \bibnamefont {Jensen}},\ }\bibfield  {title} {\bibinfo {title} {A tale of two rhythms: {{Locked}} clocks and chaos in biology},\ }\href {https://doi.org/10.1016/j.cels.2021.03.003} {\bibfield  {journal} {\bibinfo  {journal} {Cell Systems}\ }\textbf {\bibinfo {volume} {12}},\ \bibinfo {pages} {291} (\bibinfo {year} {2021})}\BibitemShut {NoStop}%
\bibitem [{\citenamefont {Matsumoto}(1984)}]{MatsumotoIEEETrans.CircuitsSyst.1984}%
  \BibitemOpen
  \bibfield  {author} {\bibinfo {author} {\bibfnamefont {T.}~\bibnamefont {Matsumoto}},\ }\bibfield  {title} {\bibinfo {title} {A chaotic attractor from {{Chua}}'s circuit},\ }\href {https://doi.org/10.1109/TCS.1984.1085459} {\bibfield  {journal} {\bibinfo  {journal} {IEEE Transactions on Circuits and Systems}\ }\textbf {\bibinfo {volume} {31}},\ \bibinfo {pages} {1055} (\bibinfo {year} {1984})}\BibitemShut {NoStop}%
\bibitem [{\citenamefont {Lorenz}(1963)}]{LorenzJ.AtmosphericSci.1963}%
  \BibitemOpen
  \bibfield  {author} {\bibinfo {author} {\bibfnamefont {E.~N.}\ \bibnamefont {Lorenz}},\ }\bibfield  {title} {\bibinfo {title} {Deterministic {{Nonperiodic Flow}}},\ }\href {https://doi.org/10.1175/1520-0469(1963)020<0130:DNF>2.0.CO;2} {\bibfield  {journal} {\bibinfo  {journal} {Journal of the Atmospheric Sciences}\ }\textbf {\bibinfo {volume} {20}},\ \bibinfo {pages} {130} (\bibinfo {year} {1963})}\BibitemShut {NoStop}%
\bibitem [{\citenamefont {Ryan}\ \emph {et~al.}(1994)\citenamefont {Ryan}, \citenamefont {Agrawal}, \citenamefont {Gray},\ and\ \citenamefont {Gage}}]{RyanIEEEJ.QuantumElectron.1994}%
  \BibitemOpen
  \bibfield  {author} {\bibinfo {author} {\bibfnamefont {A.}~\bibnamefont {Ryan}}, \bibinfo {author} {\bibfnamefont {G.}~\bibnamefont {Agrawal}}, \bibinfo {author} {\bibfnamefont {G.}~\bibnamefont {Gray}},\ and\ \bibinfo {author} {\bibfnamefont {E.}~\bibnamefont {Gage}},\ }\bibfield  {title} {\bibinfo {title} {Optical-feedback-induced chaos and its control in multimode semiconductor lasers},\ }\href {https://doi.org/10.1109/3.286153} {\bibfield  {journal} {\bibinfo  {journal} {IEEE Journal of Quantum Electronics}\ }\textbf {\bibinfo {volume} {30}},\ \bibinfo {pages} {668} (\bibinfo {year} {1994})}\BibitemShut {NoStop}%
\bibitem [{\citenamefont {Akhmediev}\ \emph {et~al.}(2001)\citenamefont {Akhmediev}, \citenamefont {{Soto-Crespo}},\ and\ \citenamefont {Town}}]{AkhmedievPhys.Rev.E2001}%
  \BibitemOpen
  \bibfield  {author} {\bibinfo {author} {\bibfnamefont {N.}~\bibnamefont {Akhmediev}}, \bibinfo {author} {\bibfnamefont {J.~M.}\ \bibnamefont {{Soto-Crespo}}},\ and\ \bibinfo {author} {\bibfnamefont {G.}~\bibnamefont {Town}},\ }\bibfield  {title} {\bibinfo {title} {Pulsating solitons, chaotic solitons, period doubling, and pulse coexistence in mode-locked lasers: {{Complex Ginzburg-Landau}} equation approach},\ }\href {https://doi.org/10.1103/PhysRevE.63.056602} {\bibfield  {journal} {\bibinfo  {journal} {Physical Review E}\ }\textbf {\bibinfo {volume} {63}},\ \bibinfo {pages} {056602} (\bibinfo {year} {2001})}\BibitemShut {NoStop}%
\bibitem [{\citenamefont {Reidler}\ \emph {et~al.}(2009)\citenamefont {Reidler}, \citenamefont {Aviad}, \citenamefont {Rosenbluh},\ and\ \citenamefont {Kanter}}]{ReidlerPhys.Rev.Lett.2009}%
  \BibitemOpen
  \bibfield  {author} {\bibinfo {author} {\bibfnamefont {I.}~\bibnamefont {Reidler}}, \bibinfo {author} {\bibfnamefont {Y.}~\bibnamefont {Aviad}}, \bibinfo {author} {\bibfnamefont {M.}~\bibnamefont {Rosenbluh}},\ and\ \bibinfo {author} {\bibfnamefont {I.}~\bibnamefont {Kanter}},\ }\bibfield  {title} {\bibinfo {title} {Ultrahigh-{{Speed Random Number Generation Based}} on a {{Chaotic Semiconductor Laser}}},\ }\href {https://doi.org/10.1103/PhysRevLett.103.024102} {\bibfield  {journal} {\bibinfo  {journal} {Physical Review Letters}\ }\textbf {\bibinfo {volume} {103}},\ \bibinfo {pages} {024102} (\bibinfo {year} {2009})}\BibitemShut {NoStop}%
\bibitem [{\citenamefont {Argyris}\ \emph {et~al.}(2005)\citenamefont {Argyris}, \citenamefont {Syvridis}, \citenamefont {Larger}, \citenamefont {{Annovazzi-Lodi}}, \citenamefont {Colet}, \citenamefont {Fischer}, \citenamefont {{Garc{\'i}a-Ojalvo}}, \citenamefont {Mirasso}, \citenamefont {Pesquera},\ and\ \citenamefont {Shore}}]{ArgyrisNature2005}%
  \BibitemOpen
  \bibfield  {author} {\bibinfo {author} {\bibfnamefont {A.}~\bibnamefont {Argyris}}, \bibinfo {author} {\bibfnamefont {D.}~\bibnamefont {Syvridis}}, \bibinfo {author} {\bibfnamefont {L.}~\bibnamefont {Larger}}, \bibinfo {author} {\bibfnamefont {V.}~\bibnamefont {{Annovazzi-Lodi}}}, \bibinfo {author} {\bibfnamefont {P.}~\bibnamefont {Colet}}, \bibinfo {author} {\bibfnamefont {I.}~\bibnamefont {Fischer}}, \bibinfo {author} {\bibfnamefont {J.}~\bibnamefont {{Garc{\'i}a-Ojalvo}}}, \bibinfo {author} {\bibfnamefont {C.~R.}\ \bibnamefont {Mirasso}}, \bibinfo {author} {\bibfnamefont {L.}~\bibnamefont {Pesquera}},\ and\ \bibinfo {author} {\bibfnamefont {K.~A.}\ \bibnamefont {Shore}},\ }\bibfield  {title} {\bibinfo {title} {Chaos-based communications at high bit rates using commercial fibre-optic links},\ }\href {https://doi.org/10.1038/nature04275} {\bibfield  {journal} {\bibinfo  {journal} {Nature}\ }\textbf {\bibinfo {volume} {438}},\ \bibinfo {pages} {343} (\bibinfo {year} {2005})}\BibitemShut {NoStop}%
\bibitem [{\citenamefont {Rafayelyan}\ \emph {et~al.}(2020)\citenamefont {Rafayelyan}, \citenamefont {Dong}, \citenamefont {Tan}, \citenamefont {Krzakala},\ and\ \citenamefont {Gigan}}]{RafayelyanPhys.Rev.X2020}%
  \BibitemOpen
  \bibfield  {author} {\bibinfo {author} {\bibfnamefont {M.}~\bibnamefont {Rafayelyan}}, \bibinfo {author} {\bibfnamefont {J.}~\bibnamefont {Dong}}, \bibinfo {author} {\bibfnamefont {Y.}~\bibnamefont {Tan}}, \bibinfo {author} {\bibfnamefont {F.}~\bibnamefont {Krzakala}},\ and\ \bibinfo {author} {\bibfnamefont {S.}~\bibnamefont {Gigan}},\ }\bibfield  {title} {\bibinfo {title} {Large-{{Scale Optical Reservoir Computing}} for {{Spatiotemporal Chaotic Systems Prediction}}},\ }\href {https://doi.org/10.1103/PhysRevX.10.041037} {\bibfield  {journal} {\bibinfo  {journal} {Physical Review X}\ }\textbf {\bibinfo {volume} {10}},\ \bibinfo {pages} {041037} (\bibinfo {year} {2020})}\BibitemShut {NoStop}%
\bibitem [{\citenamefont {Lukashchuk}\ \emph {et~al.}(2023)\citenamefont {Lukashchuk}, \citenamefont {Riemensberger}, \citenamefont {Stroganov}, \citenamefont {Navickaite},\ and\ \citenamefont {Kippenberg}}]{LukashchukAPLPhotonics2023}%
  \BibitemOpen
  \bibfield  {author} {\bibinfo {author} {\bibfnamefont {A.}~\bibnamefont {Lukashchuk}}, \bibinfo {author} {\bibfnamefont {J.}~\bibnamefont {Riemensberger}}, \bibinfo {author} {\bibfnamefont {A.}~\bibnamefont {Stroganov}}, \bibinfo {author} {\bibfnamefont {G.}~\bibnamefont {Navickaite}},\ and\ \bibinfo {author} {\bibfnamefont {T.~J.}\ \bibnamefont {Kippenberg}},\ }\bibfield  {title} {\bibinfo {title} {Chaotic microcomb inertia-free parallel ranging},\ }\href {https://doi.org/10.1063/5.0141384} {\bibfield  {journal} {\bibinfo  {journal} {APL Photonics}\ }\textbf {\bibinfo {volume} {8}},\ \bibinfo {pages} {056102} (\bibinfo {year} {2023})}\BibitemShut {NoStop}%
\bibitem [{\citenamefont {Moreno}\ \emph {et~al.}(2024)\citenamefont {Moreno}, \citenamefont {Fujii}, \citenamefont {Nakashima}, \citenamefont {Lemcke}, \citenamefont {Uchida}, \citenamefont {Sanchis},\ and\ \citenamefont {Tanabe}}]{MorenoOpt.ExpressOE2024}%
  \BibitemOpen
  \bibfield  {author} {\bibinfo {author} {\bibfnamefont {D.}~\bibnamefont {Moreno}}, \bibinfo {author} {\bibfnamefont {S.}~\bibnamefont {Fujii}}, \bibinfo {author} {\bibfnamefont {A.}~\bibnamefont {Nakashima}}, \bibinfo {author} {\bibfnamefont {D.}~\bibnamefont {Lemcke}}, \bibinfo {author} {\bibfnamefont {A.}~\bibnamefont {Uchida}}, \bibinfo {author} {\bibfnamefont {P.}~\bibnamefont {Sanchis}},\ and\ \bibinfo {author} {\bibfnamefont {T.}~\bibnamefont {Tanabe}},\ }\bibfield  {title} {\bibinfo {title} {Synchronization of two chaotic microresonator frequency combs},\ }\href {https://doi.org/10.1364/OE.511097} {\bibfield  {journal} {\bibinfo  {journal} {Optics Express}\ }\textbf {\bibinfo {volume} {32}},\ \bibinfo {pages} {2460} (\bibinfo {year} {2024})}\BibitemShut {NoStop}%
\bibitem [{\citenamefont {Kippenberg}\ \emph {et~al.}(2018)\citenamefont {Kippenberg}, \citenamefont {Gaeta}, \citenamefont {Lipson},\ and\ \citenamefont {Gorodetsky}}]{KippenbergScience2018}%
  \BibitemOpen
  \bibfield  {author} {\bibinfo {author} {\bibfnamefont {T.~J.}\ \bibnamefont {Kippenberg}}, \bibinfo {author} {\bibfnamefont {A.~L.}\ \bibnamefont {Gaeta}}, \bibinfo {author} {\bibfnamefont {M.}~\bibnamefont {Lipson}},\ and\ \bibinfo {author} {\bibfnamefont {M.~L.}\ \bibnamefont {Gorodetsky}},\ }\bibfield  {title} {\bibinfo {title} {Dissipative {{Kerr}} solitons in optical microresonators},\ }\href {https://doi.org/10.1126/science.aan8083} {\bibfield  {journal} {\bibinfo  {journal} {Science}\ }\textbf {\bibinfo {volume} {361}},\ \bibinfo {pages} {eaan8083} (\bibinfo {year} {2018})}\BibitemShut {NoStop}%
\bibitem [{\citenamefont {Chembo}\ and\ \citenamefont {Menyuk}(2013)}]{ChemboPhys.Rev.A2013}%
  \BibitemOpen
  \bibfield  {author} {\bibinfo {author} {\bibfnamefont {Y.~K.}\ \bibnamefont {Chembo}}\ and\ \bibinfo {author} {\bibfnamefont {C.~R.}\ \bibnamefont {Menyuk}},\ }\bibfield  {title} {\bibinfo {title} {Spatiotemporal {{Lugiato-Lefever}} formalism for {{Kerr-comb}} generation in whispering-gallery-mode resonators},\ }\href {https://doi.org/10.1103/PhysRevA.87.053852} {\bibfield  {journal} {\bibinfo  {journal} {Physical Review A}\ }\textbf {\bibinfo {volume} {87}},\ \bibinfo {pages} {053852} (\bibinfo {year} {2013})}\BibitemShut {NoStop}%
\bibitem [{\citenamefont {Diddams}\ \emph {et~al.}(2020)\citenamefont {Diddams}, \citenamefont {Vahala},\ and\ \citenamefont {Udem}}]{DiddamsScience2020a}%
  \BibitemOpen
  \bibfield  {author} {\bibinfo {author} {\bibfnamefont {S.~A.}\ \bibnamefont {Diddams}}, \bibinfo {author} {\bibfnamefont {K.}~\bibnamefont {Vahala}},\ and\ \bibinfo {author} {\bibfnamefont {T.}~\bibnamefont {Udem}},\ }\bibfield  {title} {\bibinfo {title} {Optical frequency combs: {{Coherently}} uniting the electromagnetic spectrum},\ }\href {https://doi.org/10.1126/science.aay3676} {\bibfield  {journal} {\bibinfo  {journal} {Science}\ }\textbf {\bibinfo {volume} {369}},\ \bibinfo {pages} {eaay3676} (\bibinfo {year} {2020})}\BibitemShut {NoStop}%
\bibitem [{\citenamefont {Grelu}\ and\ \citenamefont {Akhmediev}(2012)}]{GreluNaturePhoton2012}%
  \BibitemOpen
  \bibfield  {author} {\bibinfo {author} {\bibfnamefont {P.}~\bibnamefont {Grelu}}\ and\ \bibinfo {author} {\bibfnamefont {N.}~\bibnamefont {Akhmediev}},\ }\bibfield  {title} {\bibinfo {title} {Dissipative solitons for mode-locked lasers},\ }\href {https://doi.org/10.1038/nphoton.2011.345} {\bibfield  {journal} {\bibinfo  {journal} {Nature Photonics}\ }\textbf {\bibinfo {volume} {6}},\ \bibinfo {pages} {84} (\bibinfo {year} {2012})}\BibitemShut {NoStop}%
\bibitem [{\citenamefont {Lecaplain}\ \emph {et~al.}(2012)\citenamefont {Lecaplain}, \citenamefont {Grelu}, \citenamefont {{Soto-Crespo}},\ and\ \citenamefont {Akhmediev}}]{LecaplainPhys.Rev.Lett.2012}%
  \BibitemOpen
  \bibfield  {author} {\bibinfo {author} {\bibfnamefont {C.}~\bibnamefont {Lecaplain}}, \bibinfo {author} {\bibfnamefont {{\relax Ph}.}~\bibnamefont {Grelu}}, \bibinfo {author} {\bibfnamefont {J.~M.}\ \bibnamefont {{Soto-Crespo}}},\ and\ \bibinfo {author} {\bibfnamefont {N.}~\bibnamefont {Akhmediev}},\ }\bibfield  {title} {\bibinfo {title} {Dissipative {{Rogue Waves Generated}} by {{Chaotic Pulse Bunching}} in a {{Mode-Locked Laser}}},\ }\href {https://doi.org/10.1103/PhysRevLett.108.233901} {\bibfield  {journal} {\bibinfo  {journal} {Physical Review Letters}\ }\textbf {\bibinfo {volume} {108}},\ \bibinfo {pages} {233901} (\bibinfo {year} {2012})}\BibitemShut {NoStop}%
\bibitem [{\citenamefont {Song}\ \emph {et~al.}(2023)\citenamefont {Song}, \citenamefont {Zou}, \citenamefont {Gat}, \citenamefont {Hu},\ and\ \citenamefont {Grelu}}]{SongLaserPhotonicsRev.2023}%
  \BibitemOpen
  \bibfield  {author} {\bibinfo {author} {\bibfnamefont {Y.}~\bibnamefont {Song}}, \bibinfo {author} {\bibfnamefont {D.}~\bibnamefont {Zou}}, \bibinfo {author} {\bibfnamefont {O.}~\bibnamefont {Gat}}, \bibinfo {author} {\bibfnamefont {M.}~\bibnamefont {Hu}},\ and\ \bibinfo {author} {\bibfnamefont {P.}~\bibnamefont {Grelu}},\ }\bibfield  {title} {\bibinfo {title} {Chaotic {{Internal Dynamics}} of {{Dissipative Optical Soliton Molecules}}},\ }\href {https://doi.org/10.1002/lpor.202300066} {\bibfield  {journal} {\bibinfo  {journal} {Laser \& Photonics Reviews}\ }\textbf {\bibinfo {volume} {17}},\ \bibinfo {pages} {2300066} (\bibinfo {year} {2023})}\BibitemShut {NoStop}%
\bibitem [{\citenamefont {Moille}\ \emph {et~al.}(2023)\citenamefont {Moille}, \citenamefont {Stone}, \citenamefont {Chojnacky}, \citenamefont {Shrestha}, \citenamefont {Javid}, \citenamefont {Menyuk},\ and\ \citenamefont {Srinivasan}}]{MoilleNature2023}%
  \BibitemOpen
  \bibfield  {author} {\bibinfo {author} {\bibfnamefont {G.}~\bibnamefont {Moille}}, \bibinfo {author} {\bibfnamefont {J.}~\bibnamefont {Stone}}, \bibinfo {author} {\bibfnamefont {M.}~\bibnamefont {Chojnacky}}, \bibinfo {author} {\bibfnamefont {R.}~\bibnamefont {Shrestha}}, \bibinfo {author} {\bibfnamefont {U.~A.}\ \bibnamefont {Javid}}, \bibinfo {author} {\bibfnamefont {C.}~\bibnamefont {Menyuk}},\ and\ \bibinfo {author} {\bibfnamefont {K.}~\bibnamefont {Srinivasan}},\ }\bibfield  {title} {\bibinfo {title} {Kerr-induced synchronization of a cavity soliton to an optical reference},\ }\href {https://doi.org/10.1038/s41586-023-06730-0} {\bibfield  {journal} {\bibinfo  {journal} {Nature}\ }\textbf {\bibinfo {volume} {624}},\ \bibinfo {pages} {267} (\bibinfo {year} {2023})}\BibitemShut {NoStop}%
\bibitem [{\citenamefont {Wildi}\ \emph {et~al.}(2023)\citenamefont {Wildi}, \citenamefont {Ulanov}, \citenamefont {Englebert}, \citenamefont {Voumard},\ and\ \citenamefont {Herr}}]{WildiAPLPhotonics2023}%
  \BibitemOpen
  \bibfield  {author} {\bibinfo {author} {\bibfnamefont {T.}~\bibnamefont {Wildi}}, \bibinfo {author} {\bibfnamefont {A.}~\bibnamefont {Ulanov}}, \bibinfo {author} {\bibfnamefont {N.}~\bibnamefont {Englebert}}, \bibinfo {author} {\bibfnamefont {T.}~\bibnamefont {Voumard}},\ and\ \bibinfo {author} {\bibfnamefont {T.}~\bibnamefont {Herr}},\ }\bibfield  {title} {\bibinfo {title} {Sideband injection locking in microresonator frequency combs},\ }\href {https://doi.org/10.1063/5.0170224} {\bibfield  {journal} {\bibinfo  {journal} {APL Photonics}\ }\textbf {\bibinfo {volume} {8}},\ \bibinfo {pages} {120801} (\bibinfo {year} {2023})}\BibitemShut {NoStop}%
\bibitem [{\citenamefont {Moille}\ \emph {et~al.}(2025{\natexlab{a}})\citenamefont {Moille}, \citenamefont {Shandilya}, \citenamefont {Stone}, \citenamefont {Menyuk},\ and\ \citenamefont {Srinivasan}}]{Moille2024_arXivTRN}%
  \BibitemOpen
  \bibfield  {author} {\bibinfo {author} {\bibfnamefont {G.}~\bibnamefont {Moille}}, \bibinfo {author} {\bibfnamefont {P.}~\bibnamefont {Shandilya}}, \bibinfo {author} {\bibfnamefont {J.}~\bibnamefont {Stone}}, \bibinfo {author} {\bibfnamefont {C.}~\bibnamefont {Menyuk}},\ and\ \bibinfo {author} {\bibfnamefont {K.}~\bibnamefont {Srinivasan}},\ }\bibfield  {title} {\bibinfo {title} {All-optical noise quenching of an integrated frequency comb},\ }\href {https://doi.org/10.1364/OPTICA.561954} {\bibfield  {journal} {\bibinfo  {journal} {Optica}\ }\textbf {\bibinfo {volume} {12}},\ \bibinfo {pages} {1020} (\bibinfo {year} {2025}{\natexlab{a}})}\BibitemShut {NoStop}%
\bibitem [{\citenamefont {Jang}\ \emph {et~al.}(2015)\citenamefont {Jang}, \citenamefont {Erkintalo}, \citenamefont {Coen},\ and\ \citenamefont {Murdoch}}]{JangNatCommun2015a}%
  \BibitemOpen
  \bibfield  {author} {\bibinfo {author} {\bibfnamefont {J.~K.}\ \bibnamefont {Jang}}, \bibinfo {author} {\bibfnamefont {M.}~\bibnamefont {Erkintalo}}, \bibinfo {author} {\bibfnamefont {S.}~\bibnamefont {Coen}},\ and\ \bibinfo {author} {\bibfnamefont {S.~G.}\ \bibnamefont {Murdoch}},\ }\bibfield  {title} {\bibinfo {title} {Temporal tweezing of light through the trapping and manipulation of temporal cavity solitons},\ }\href {https://doi.org/10.1038/ncomms8370} {\bibfield  {journal} {\bibinfo  {journal} {Nature Communications}\ }\textbf {\bibinfo {volume} {6}},\ \bibinfo {pages} {7370} (\bibinfo {year} {2015})}\BibitemShut {NoStop}%
\bibitem [{\citenamefont {Jang}\ \emph {et~al.}(2018)\citenamefont {Jang}, \citenamefont {Klenner}, \citenamefont {Ji}, \citenamefont {Okawachi}, \citenamefont {Lipson},\ and\ \citenamefont {Gaeta}}]{JangNaturePhoton2018b}%
  \BibitemOpen
  \bibfield  {author} {\bibinfo {author} {\bibfnamefont {J.~K.}\ \bibnamefont {Jang}}, \bibinfo {author} {\bibfnamefont {A.}~\bibnamefont {Klenner}}, \bibinfo {author} {\bibfnamefont {X.}~\bibnamefont {Ji}}, \bibinfo {author} {\bibfnamefont {Y.}~\bibnamefont {Okawachi}}, \bibinfo {author} {\bibfnamefont {M.}~\bibnamefont {Lipson}},\ and\ \bibinfo {author} {\bibfnamefont {A.~L.}\ \bibnamefont {Gaeta}},\ }\bibfield  {title} {\bibinfo {title} {Synchronization of coupled optical microresonators},\ }\href {https://doi.org/10.1038/s41566-018-0261-x} {\bibfield  {journal} {\bibinfo  {journal} {Nature Photonics}\ }\textbf {\bibinfo {volume} {12}},\ \bibinfo {pages} {688} (\bibinfo {year} {2018})}\BibitemShut {NoStop}%
\bibitem [{\citenamefont {Englebert}\ \emph {et~al.}(2023)\citenamefont {Englebert}, \citenamefont {Goldman}, \citenamefont {Erkintalo}, \citenamefont {Mostaan}, \citenamefont {Gorza}, \citenamefont {Leo},\ and\ \citenamefont {Fatome}}]{EnglebertNat.Phys.2023}%
  \BibitemOpen
  \bibfield  {author} {\bibinfo {author} {\bibfnamefont {N.}~\bibnamefont {Englebert}}, \bibinfo {author} {\bibfnamefont {N.}~\bibnamefont {Goldman}}, \bibinfo {author} {\bibfnamefont {M.}~\bibnamefont {Erkintalo}}, \bibinfo {author} {\bibfnamefont {N.}~\bibnamefont {Mostaan}}, \bibinfo {author} {\bibfnamefont {S.-P.}\ \bibnamefont {Gorza}}, \bibinfo {author} {\bibfnamefont {F.}~\bibnamefont {Leo}},\ and\ \bibinfo {author} {\bibfnamefont {J.}~\bibnamefont {Fatome}},\ }\bibfield  {title} {\bibinfo {title} {Bloch oscillations of coherently driven dissipative solitons in a synthetic dimension},\ }\href {https://doi.org/10.1038/s41567-023-02005-7} {\bibfield  {journal} {\bibinfo  {journal} {Nature Physics}\ }\textbf {\bibinfo {volume} {19}},\ \bibinfo {pages} {1014} (\bibinfo {year} {2023})}\BibitemShut {NoStop}%
\bibitem [{\citenamefont {Cole}\ and\ \citenamefont {Papp}(2019)}]{ColePhys.Rev.Lett.2019}%
  \BibitemOpen
  \bibfield  {author} {\bibinfo {author} {\bibfnamefont {D.~C.}\ \bibnamefont {Cole}}\ and\ \bibinfo {author} {\bibfnamefont {S.~B.}\ \bibnamefont {Papp}},\ }\bibfield  {title} {\bibinfo {title} {Subharmonic {{Entrainment}} of {{Kerr Breather Solitons}}},\ }\href {https://doi.org/10.1103/PhysRevLett.123.173904} {\bibfield  {journal} {\bibinfo  {journal} {Physical Review Letters}\ }\textbf {\bibinfo {volume} {123}},\ \bibinfo {pages} {173904} (\bibinfo {year} {2019})}\BibitemShut {NoStop}%
\bibitem [{\citenamefont {Yang}\ \emph {et~al.}(2017)\citenamefont {Yang}, \citenamefont {Yi}, \citenamefont {Yang},\ and\ \citenamefont {Vahala}}]{YangNaturePhoton2017}%
  \BibitemOpen
  \bibfield  {author} {\bibinfo {author} {\bibfnamefont {Q.-F.}\ \bibnamefont {Yang}}, \bibinfo {author} {\bibfnamefont {X.}~\bibnamefont {Yi}}, \bibinfo {author} {\bibfnamefont {K.~Y.}\ \bibnamefont {Yang}},\ and\ \bibinfo {author} {\bibfnamefont {K.}~\bibnamefont {Vahala}},\ }\bibfield  {title} {\bibinfo {title} {Counter-propagating solitons in microresonators},\ }\href {https://doi.org/10.1038/nphoton.2017.117} {\bibfield  {journal} {\bibinfo  {journal} {Nature Photonics}\ }\textbf {\bibinfo {volume} {11}},\ \bibinfo {pages} {560} (\bibinfo {year} {2017})}\BibitemShut {NoStop}%
\bibitem [{\citenamefont {Sun}\ \emph {et~al.}(2025)\citenamefont {Sun}, \citenamefont {Harrington}, \citenamefont {Tabatabaei}, \citenamefont {Hanifi}, \citenamefont {Liu}, \citenamefont {Wang}, \citenamefont {Wang}, \citenamefont {Yang}, \citenamefont {Liu}, \citenamefont {Morgan}, \citenamefont {Bowers}, \citenamefont {Morton}, \citenamefont {Nelson}, \citenamefont {Beling}, \citenamefont {Blumenthal},\ and\ \citenamefont {Yi}}]{Sun2024}%
  \BibitemOpen
  \bibfield  {author} {\bibinfo {author} {\bibfnamefont {S.}~\bibnamefont {Sun}}, \bibinfo {author} {\bibfnamefont {M.~W.}\ \bibnamefont {Harrington}}, \bibinfo {author} {\bibfnamefont {F.}~\bibnamefont {Tabatabaei}}, \bibinfo {author} {\bibfnamefont {S.}~\bibnamefont {Hanifi}}, \bibinfo {author} {\bibfnamefont {K.}~\bibnamefont {Liu}}, \bibinfo {author} {\bibfnamefont {J.}~\bibnamefont {Wang}}, \bibinfo {author} {\bibfnamefont {B.}~\bibnamefont {Wang}}, \bibinfo {author} {\bibfnamefont {Z.}~\bibnamefont {Yang}}, \bibinfo {author} {\bibfnamefont {R.}~\bibnamefont {Liu}}, \bibinfo {author} {\bibfnamefont {J.~S.}\ \bibnamefont {Morgan}}, \bibinfo {author} {\bibfnamefont {S.~M.}\ \bibnamefont {Bowers}}, \bibinfo {author} {\bibfnamefont {P.~A.}\ \bibnamefont {Morton}}, \bibinfo {author} {\bibfnamefont {K.~D.}\ \bibnamefont {Nelson}}, \bibinfo {author} {\bibfnamefont {A.}~\bibnamefont {Beling}}, \bibinfo {author} {\bibfnamefont {D.~J.}\ \bibnamefont {Blumenthal}},\ and\ \bibinfo {author} {\bibfnamefont {X.}~\bibnamefont {Yi}},\ }\bibfield  {title} {\bibinfo {title} {Microcavity {{Kerr}} optical frequency division with integrated {{SiN}} photonics},\ }\href {https://doi.org/10.1038/s41566-025-01668-3} {\bibfield  {journal} {\bibinfo  {journal} {Nature Photonics}\ ,\ \bibinfo {pages} {1}} (\bibinfo {year} {2025})}\BibitemShut {NoStop}%
\bibitem [{\citenamefont {Valizadeh}\ \emph {et~al.}(2008)\citenamefont {Valizadeh}, \citenamefont {Kolahchi},\ and\ \citenamefont {Straley}}]{ValizadehJNMP2008}%
  \BibitemOpen
  \bibfield  {author} {\bibinfo {author} {\bibfnamefont {A.}~\bibnamefont {Valizadeh}}, \bibinfo {author} {\bibfnamefont {M.}~\bibnamefont {Kolahchi}},\ and\ \bibinfo {author} {\bibfnamefont {J.}~\bibnamefont {Straley}},\ }\bibfield  {title} {\bibinfo {title} {On the {{Origin}} of {{Fractional Shapiro Steps}} in {{Systems}} of {{Josephson Junctions}} with {{Few Degrees}} of {{Freedom}}:},\ }\href {https://doi.org/10.2991/jnmp.2008.15.s3.39} {\bibfield  {journal} {\bibinfo  {journal} {Journal of Nonlinear Mathematical Physics}\ }\textbf {\bibinfo {volume} {15}},\ \bibinfo {pages} {407} (\bibinfo {year} {2008})}\BibitemShut {NoStop}%
\bibitem [{\citenamefont {Taheri}\ \emph {et~al.}(2017)\citenamefont {Taheri}, \citenamefont {Matsko},\ and\ \citenamefont {Maleki}}]{TaheriEur.Phys.J.D2017}%
  \BibitemOpen
  \bibfield  {author} {\bibinfo {author} {\bibfnamefont {H.}~\bibnamefont {Taheri}}, \bibinfo {author} {\bibfnamefont {A.~B.}\ \bibnamefont {Matsko}},\ and\ \bibinfo {author} {\bibfnamefont {L.}~\bibnamefont {Maleki}},\ }\bibfield  {title} {\bibinfo {title} {Optical lattice trap for {{Kerr}} solitons},\ }\href {https://doi.org/10.1140/epjd/e2017-80150-6} {\bibfield  {journal} {\bibinfo  {journal} {The European Physical Journal D}\ }\textbf {\bibinfo {volume} {71}},\ \bibinfo {pages} {153} (\bibinfo {year} {2017})}\BibitemShut {NoStop}%
\bibitem [{\citenamefont {Coullet}\ \emph {et~al.}(2005)\citenamefont {Coullet}, \citenamefont {Gilli}, \citenamefont {Monticelli},\ and\ \citenamefont {Vandenberghe}}]{CoulletAmericanJournalofPhysics2005}%
  \BibitemOpen
  \bibfield  {author} {\bibinfo {author} {\bibfnamefont {P.}~\bibnamefont {Coullet}}, \bibinfo {author} {\bibfnamefont {J.~M.}\ \bibnamefont {Gilli}}, \bibinfo {author} {\bibfnamefont {M.}~\bibnamefont {Monticelli}},\ and\ \bibinfo {author} {\bibfnamefont {N.}~\bibnamefont {Vandenberghe}},\ }\bibfield  {title} {\bibinfo {title} {A damped pendulum forced with a constant torque},\ }\href {https://doi.org/10.1119/1.2074027} {\bibfield  {journal} {\bibinfo  {journal} {American Journal of Physics}\ }\textbf {\bibinfo {volume} {73}},\ \bibinfo {pages} {1122} (\bibinfo {year} {2005})}\BibitemShut {NoStop}%
\bibitem [{\citenamefont {Moille}\ \emph {et~al.}(2019{\natexlab{a}})\citenamefont {Moille}, \citenamefont {Li}, \citenamefont {Briles}, \citenamefont {Yu}, \citenamefont {Drake}, \citenamefont {Lu}, \citenamefont {Rao}, \citenamefont {Westly}, \citenamefont {Papp},\ and\ \citenamefont {Srinivasan}}]{MoilleOpt.Lett.2019}%
  \BibitemOpen
  \bibfield  {author} {\bibinfo {author} {\bibfnamefont {G.}~\bibnamefont {Moille}}, \bibinfo {author} {\bibfnamefont {Q.}~\bibnamefont {Li}}, \bibinfo {author} {\bibfnamefont {T.~C.}\ \bibnamefont {Briles}}, \bibinfo {author} {\bibfnamefont {S.-P.}\ \bibnamefont {Yu}}, \bibinfo {author} {\bibfnamefont {T.}~\bibnamefont {Drake}}, \bibinfo {author} {\bibfnamefont {X.}~\bibnamefont {Lu}}, \bibinfo {author} {\bibfnamefont {A.}~\bibnamefont {Rao}}, \bibinfo {author} {\bibfnamefont {D.}~\bibnamefont {Westly}}, \bibinfo {author} {\bibfnamefont {S.~B.}\ \bibnamefont {Papp}},\ and\ \bibinfo {author} {\bibfnamefont {K.}~\bibnamefont {Srinivasan}},\ }\bibfield  {title} {\bibinfo {title} {Broadband resonator-waveguide coupling for efficient extraction of octave-spanning microcombs},\ }\href {https://doi.org/10.1364/OL.44.004737} {\bibfield  {journal} {\bibinfo  {journal} {Optics Letters}\ }\textbf {\bibinfo {volume} {44}},\ \bibinfo {pages} {4737} (\bibinfo {year} {2019}{\natexlab{a}})}\BibitemShut {NoStop}%
\bibitem [{\citenamefont {Zhang}\ \emph {et~al.}(2019)\citenamefont {Zhang}, \citenamefont {Silver}, \citenamefont {Del~Bino}, \citenamefont {Copie}, \citenamefont {Woodley}, \citenamefont {Ghalanos}, \citenamefont {Svela}, \citenamefont {Moroney},\ and\ \citenamefont {Del'Haye}}]{ZhangOptica2019}%
  \BibitemOpen
  \bibfield  {author} {\bibinfo {author} {\bibfnamefont {S.}~\bibnamefont {Zhang}}, \bibinfo {author} {\bibfnamefont {J.~M.}\ \bibnamefont {Silver}}, \bibinfo {author} {\bibfnamefont {L.}~\bibnamefont {Del~Bino}}, \bibinfo {author} {\bibfnamefont {F.}~\bibnamefont {Copie}}, \bibinfo {author} {\bibfnamefont {M.~T.~M.}\ \bibnamefont {Woodley}}, \bibinfo {author} {\bibfnamefont {G.~N.}\ \bibnamefont {Ghalanos}}, \bibinfo {author} {\bibfnamefont {A.~{\O}.}\ \bibnamefont {Svela}}, \bibinfo {author} {\bibfnamefont {N.}~\bibnamefont {Moroney}},\ and\ \bibinfo {author} {\bibfnamefont {P.}~\bibnamefont {Del'Haye}},\ }\bibfield  {title} {\bibinfo {title} {Sub-milliwatt-level microresonator solitons with extended access range using an auxiliary laser},\ }\href {https://doi.org/10.1364/OPTICA.6.000206} {\bibfield  {journal} {\bibinfo  {journal} {Optica}\ }\textbf {\bibinfo {volume} {6}},\ \bibinfo {pages} {206} (\bibinfo {year} {2019})}\BibitemShut {NoStop}%
\bibitem [{\citenamefont {Zhou}\ \emph {et~al.}(2019)\citenamefont {Zhou}, \citenamefont {Geng}, \citenamefont {Cui}, \citenamefont {Huang}, \citenamefont {Zhou}, \citenamefont {Qiu},\ and\ \citenamefont {Wei~Wong}}]{ZhouLightSciAppl2019}%
  \BibitemOpen
  \bibfield  {author} {\bibinfo {author} {\bibfnamefont {H.}~\bibnamefont {Zhou}}, \bibinfo {author} {\bibfnamefont {Y.}~\bibnamefont {Geng}}, \bibinfo {author} {\bibfnamefont {W.}~\bibnamefont {Cui}}, \bibinfo {author} {\bibfnamefont {S.-W.}\ \bibnamefont {Huang}}, \bibinfo {author} {\bibfnamefont {Q.}~\bibnamefont {Zhou}}, \bibinfo {author} {\bibfnamefont {K.}~\bibnamefont {Qiu}},\ and\ \bibinfo {author} {\bibfnamefont {C.}~\bibnamefont {Wei~Wong}},\ }\bibfield  {title} {\bibinfo {title} {Soliton bursts and deterministic dissipative {{Kerr}} soliton generation in auxiliary-assisted microcavities},\ }\href {https://doi.org/10.1038/s41377-019-0161-y} {\bibfield  {journal} {\bibinfo  {journal} {Light: Science \& Applications}\ }\textbf {\bibinfo {volume} {8}},\ \bibinfo {pages} {50} (\bibinfo {year} {2019})}\BibitemShut {NoStop}%
\bibitem [{\citenamefont {Stone}\ and\ \citenamefont {Papp}(2020)}]{StonePhys.Rev.Lett.2020}%
  \BibitemOpen
  \bibfield  {author} {\bibinfo {author} {\bibfnamefont {J.~R.}\ \bibnamefont {Stone}}\ and\ \bibinfo {author} {\bibfnamefont {S.~B.}\ \bibnamefont {Papp}},\ }\bibfield  {title} {\bibinfo {title} {Harnessing {{Dispersion}} in {{Soliton Microcombs}} to {{Mitigate Thermal Noise}}},\ }\href {https://doi.org/10.1103/PhysRevLett.125.153901} {\bibfield  {journal} {\bibinfo  {journal} {Physical Review Letters}\ }\textbf {\bibinfo {volume} {125}},\ \bibinfo {pages} {153901} (\bibinfo {year} {2020})}\BibitemShut {NoStop}%
\bibitem [{\citenamefont {Wang}\ \emph {et~al.}(2017)\citenamefont {Wang}, \citenamefont {Leo}, \citenamefont {Fatome}, \citenamefont {Erkintalo}, \citenamefont {Murdoch},\ and\ \citenamefont {Coen}}]{WangOptica2017a}%
  \BibitemOpen
  \bibfield  {author} {\bibinfo {author} {\bibfnamefont {Y.}~\bibnamefont {Wang}}, \bibinfo {author} {\bibfnamefont {F.}~\bibnamefont {Leo}}, \bibinfo {author} {\bibfnamefont {J.}~\bibnamefont {Fatome}}, \bibinfo {author} {\bibfnamefont {M.}~\bibnamefont {Erkintalo}}, \bibinfo {author} {\bibfnamefont {S.~G.}\ \bibnamefont {Murdoch}},\ and\ \bibinfo {author} {\bibfnamefont {S.}~\bibnamefont {Coen}},\ }\bibfield  {title} {\bibinfo {title} {Universal mechanism for the binding of temporal cavity solitons},\ }\href {https://doi.org/10.1364/OPTICA.4.000855} {\bibfield  {journal} {\bibinfo  {journal} {Optica}\ }\textbf {\bibinfo {volume} {4}},\ \bibinfo {pages} {855} (\bibinfo {year} {2017})}\BibitemShut {NoStop}%
\bibitem [{\citenamefont {Rukhin}\ \emph {et~al.}(2010)\citenamefont {Rukhin}, \citenamefont {Soto}, \citenamefont {Nechvatal}, \citenamefont {Smid}, \citenamefont {Barker}, \citenamefont {Leigh}, \citenamefont {Levenson}, \citenamefont {Vangel}, \citenamefont {Banks}, \citenamefont {Heckert}, \citenamefont {Dray}, \citenamefont {Vo},\ and\ \citenamefont {Bassham}}]{Rukhin2010}%
  \BibitemOpen
  \bibfield  {author} {\bibinfo {author} {\bibfnamefont {A.}~\bibnamefont {Rukhin}}, \bibinfo {author} {\bibfnamefont {J.}~\bibnamefont {Soto}}, \bibinfo {author} {\bibfnamefont {J.}~\bibnamefont {Nechvatal}}, \bibinfo {author} {\bibfnamefont {M.}~\bibnamefont {Smid}}, \bibinfo {author} {\bibfnamefont {E.}~\bibnamefont {Barker}}, \bibinfo {author} {\bibfnamefont {S.}~\bibnamefont {Leigh}}, \bibinfo {author} {\bibfnamefont {M.}~\bibnamefont {Levenson}}, \bibinfo {author} {\bibfnamefont {M.}~\bibnamefont {Vangel}}, \bibinfo {author} {\bibfnamefont {D.}~\bibnamefont {Banks}}, \bibinfo {author} {\bibfnamefont {N.}~\bibnamefont {Heckert}}, \bibinfo {author} {\bibfnamefont {J.}~\bibnamefont {Dray}}, \bibinfo {author} {\bibfnamefont {S.}~\bibnamefont {Vo}},\ and\ \bibinfo {author} {\bibfnamefont {L.}~\bibnamefont {Bassham}},\ }\href {https://doi.org/10.6028/NIST.SP.800-22r1a} {\emph {\bibinfo {title} {A {{Statistical Test Suite}} for {{Random}} and {{Pseudorandom Number Generators}} for {{Cryptographic Applications}}}}},\ \bibinfo {type} {Tech. Rep.}\ \bibinfo {number} {NIST Special Publication (SP) 800-22 Rev. 1}\ (\bibinfo  {institution} {{National Institute of Standards and Technology}},\ \bibinfo {year} {2010})\BibitemShut {NoStop}%
\bibitem [{\citenamefont {Ang}()}]{RandomGithub}%
  \BibitemOpen
  \bibfield  {author} {\bibinfo {author} {\bibfnamefont {S.~K.}\ \bibnamefont {Ang}},\ }\href@noop {} {\bibinfo {title} {Randomness test suit}},\ \bibinfo {howpublished} {\url{https://github.com/stevenang/randomness_testsuite}}\BibitemShut {NoStop}%
\bibitem [{\citenamefont {McCumber}(1968)}]{McCumberJ.Appl.Phys.1968}%
  \BibitemOpen
  \bibfield  {author} {\bibinfo {author} {\bibfnamefont {D.~E.}\ \bibnamefont {McCumber}},\ }\bibfield  {title} {\bibinfo {title} {Effect of ac {{Impedance}} on dc {{Voltage-Current Characteristics}} of {{Superconductor Weak-Link Junctions}}},\ }\href {https://doi.org/10.1063/1.1656743} {\bibfield  {journal} {\bibinfo  {journal} {Journal of Applied Physics}\ }\textbf {\bibinfo {volume} {39}},\ \bibinfo {pages} {3113} (\bibinfo {year} {1968})}\BibitemShut {NoStop}%
\bibitem [{\citenamefont {Philip}(1979)}]{PhilipPhil.Trans.R.Soc.Lond.A1979}%
  \BibitemOpen
  \bibfield  {author} {\bibinfo {author} {\bibfnamefont {H.}~\bibnamefont {Philip}},\ }\bibfield  {title} {\bibinfo {title} {A nonlinear oscillator with a strange attractor},\ }\href {https://doi.org/10.1098/rsta.1979.0068} {\bibfield  {journal} {\bibinfo  {journal} {Philosophical Transactions of the Royal Society of London. Series A, Mathematical and Physical Sciences}\ }\textbf {\bibinfo {volume} {292}},\ \bibinfo {pages} {419} (\bibinfo {year} {1979})}\BibitemShut {NoStop}%
\bibitem [{\citenamefont {Josephson}(1962)}]{JosephsonPhysicsLetters1962a}%
  \BibitemOpen
  \bibfield  {author} {\bibinfo {author} {\bibfnamefont {B.~D.}\ \bibnamefont {Josephson}},\ }\bibfield  {title} {\bibinfo {title} {Possible new effects in superconductive tunnelling},\ }\href {https://doi.org/10.1016/0031-9163(62)91369-0} {\bibfield  {journal} {\bibinfo  {journal} {Physics Letters}\ }\textbf {\bibinfo {volume} {1}},\ \bibinfo {pages} {251} (\bibinfo {year} {1962})}\BibitemShut {NoStop}%
\bibitem [{\citenamefont {Shapiro}(1963)}]{ShapiroPhys.Rev.Lett.1963}%
  \BibitemOpen
  \bibfield  {author} {\bibinfo {author} {\bibfnamefont {S.}~\bibnamefont {Shapiro}},\ }\bibfield  {title} {\bibinfo {title} {Josephson {{Currents}} in {{Superconducting Tunneling}}: {{The Effect}} of {{Microwaves}} and {{Other Observations}}},\ }\href {https://doi.org/10.1103/PhysRevLett.11.80} {\bibfield  {journal} {\bibinfo  {journal} {Physical Review Letters}\ }\textbf {\bibinfo {volume} {11}},\ \bibinfo {pages} {80} (\bibinfo {year} {1963})}\BibitemShut {NoStop}%
\bibitem [{\citenamefont {Menyuk}\ \emph {et~al.}(2025)\citenamefont {Menyuk}, \citenamefont {Shandilya}, \citenamefont {Courtright}, \citenamefont {Moille},\ and\ \citenamefont {Srinivasan}}]{Menyuk2024}%
  \BibitemOpen
  \bibfield  {author} {\bibinfo {author} {\bibfnamefont {C.~R.}\ \bibnamefont {Menyuk}}, \bibinfo {author} {\bibfnamefont {P.}~\bibnamefont {Shandilya}}, \bibinfo {author} {\bibfnamefont {L.}~\bibnamefont {Courtright}}, \bibinfo {author} {\bibfnamefont {G.}~\bibnamefont {Moille}},\ and\ \bibinfo {author} {\bibfnamefont {K.}~\bibnamefont {Srinivasan}},\ }\bibfield  {title} {\bibinfo {title} {Multi-color solitons and frequency combs in microresonators},\ }\href {https://doi.org/10.1364/OE.544077} {\bibfield  {journal} {\bibinfo  {journal} {Optics Express}\ }\textbf {\bibinfo {volume} {33}},\ \bibinfo {pages} {21824} (\bibinfo {year} {2025})}\BibitemShut {NoStop}%
\bibitem [{\citenamefont {Taheri}\ \emph {et~al.}(2025)\citenamefont {Taheri}, \citenamefont {Matsko},\ and\ \citenamefont {Wiesenfeld}}]{TaheriLaserPhotonicsRev.2025}%
  \BibitemOpen
  \bibfield  {author} {\bibinfo {author} {\bibfnamefont {H.}~\bibnamefont {Taheri}}, \bibinfo {author} {\bibfnamefont {A.~B.}\ \bibnamefont {Matsko}},\ and\ \bibinfo {author} {\bibfnamefont {K.~A.}\ \bibnamefont {Wiesenfeld}},\ }\bibfield  {title} {\bibinfo {title} {Dynamics of {{Dissipative Cavity Solitons}} in a {{Lattice Trap}}},\ }\href {https://doi.org/10.1002/lpor.202500257} {\bibfield  {journal} {\bibinfo  {journal} {Laser \& Photonics Reviews}\ }\textbf {\bibinfo {volume} {n/a}},\ \bibinfo {pages} {2500257} (\bibinfo {year} {2025})}\BibitemShut {NoStop}%
\bibitem [{\citenamefont {Moille}\ \emph {et~al.}(2025{\natexlab{b}})\citenamefont {Moille}, \citenamefont {Shandilya}, \citenamefont {Niang}, \citenamefont {Menyuk}, \citenamefont {Carter},\ and\ \citenamefont {Srinivasan}}]{MoilleNat.Photon.2025}%
  \BibitemOpen
  \bibfield  {author} {\bibinfo {author} {\bibfnamefont {G.}~\bibnamefont {Moille}}, \bibinfo {author} {\bibfnamefont {P.}~\bibnamefont {Shandilya}}, \bibinfo {author} {\bibfnamefont {A.}~\bibnamefont {Niang}}, \bibinfo {author} {\bibfnamefont {C.}~\bibnamefont {Menyuk}}, \bibinfo {author} {\bibfnamefont {G.}~\bibnamefont {Carter}},\ and\ \bibinfo {author} {\bibfnamefont {K.}~\bibnamefont {Srinivasan}},\ }\bibfield  {title} {\bibinfo {title} {Versatile optical frequency division with {{Kerr-induced}} synchronization at tunable microcomb synthetic dispersive waves},\ }\href {https://doi.org/10.1038/s41566-024-01540-w} {\bibfield  {journal} {\bibinfo  {journal} {Nature Photonics}\ }\textbf {\bibinfo {volume} {19}},\ \bibinfo {pages} {36} (\bibinfo {year} {2025}{\natexlab{b}})}\BibitemShut {NoStop}%
\bibitem [{\citenamefont {Moille}\ \emph {et~al.}(2025{\natexlab{c}})\citenamefont {Moille}, \citenamefont {Shandilya}, \citenamefont {Erkintalo}, \citenamefont {Menyuk},\ and\ \citenamefont {Srinivasan}}]{MoilleArXivParametric2024}%
  \BibitemOpen
  \bibfield  {author} {\bibinfo {author} {\bibfnamefont {G.}~\bibnamefont {Moille}}, \bibinfo {author} {\bibfnamefont {P.}~\bibnamefont {Shandilya}}, \bibinfo {author} {\bibfnamefont {M.}~\bibnamefont {Erkintalo}}, \bibinfo {author} {\bibfnamefont {C.~R.}\ \bibnamefont {Menyuk}},\ and\ \bibinfo {author} {\bibfnamefont {K.}~\bibnamefont {Srinivasan}},\ }\bibfield  {title} {\bibinfo {title} {On-{{Chip Parametric Synchronization}} of a {{Dissipative Kerr Soliton Microcomb}}},\ }\href {https://doi.org/10.1103/PhysRevLett.134.193802} {\bibfield  {journal} {\bibinfo  {journal} {Physical Review Letters}\ }\textbf {\bibinfo {volume} {134}},\ \bibinfo {pages} {193802} (\bibinfo {year} {2025}{\natexlab{c}})}\BibitemShut {NoStop}%
\bibitem [{\citenamefont {Strogatz}(2019)}]{Strogatz2019}%
  \BibitemOpen
  \bibfield  {author} {\bibinfo {author} {\bibfnamefont {S.}~\bibnamefont {Strogatz}},\ }\bibfield  {title} {\bibinfo {title} {Nonlinear dynamics and chaos: With applications to physics, biology, chemistry, and engineering}\ }(\bibinfo {address} {Boca Raton London New York},\ \bibinfo {year} {2019})\ \bibinfo {edition} {second edition}\ ed.,\ Chap.~\bibinfo {chapter} {8}\BibitemShut {NoStop}%
\bibitem [{\citenamefont {Wolf}\ \emph {et~al.}(1985)\citenamefont {Wolf}, \citenamefont {Swift}, \citenamefont {Swinney},\ and\ \citenamefont {Vastano}}]{WolfPhysicaD:NonlinearPhenomena1985}%
  \BibitemOpen
  \bibfield  {author} {\bibinfo {author} {\bibfnamefont {A.}~\bibnamefont {Wolf}}, \bibinfo {author} {\bibfnamefont {J.~B.}\ \bibnamefont {Swift}}, \bibinfo {author} {\bibfnamefont {H.~L.}\ \bibnamefont {Swinney}},\ and\ \bibinfo {author} {\bibfnamefont {J.~A.}\ \bibnamefont {Vastano}},\ }\bibfield  {title} {\bibinfo {title} {Determining {{Lyapunov}} exponents from a time series},\ }\href {https://doi.org/10.1016/0167-2789(85)90011-9} {\bibfield  {journal} {\bibinfo  {journal} {Physica D: Nonlinear Phenomena}\ }\textbf {\bibinfo {volume} {16}},\ \bibinfo {pages} {285} (\bibinfo {year} {1985})}\BibitemShut {NoStop}%
\bibitem [{\citenamefont {Gottwald}\ and\ \citenamefont {Melbourne}(2009)}]{GottwaldSIAMJ.Appl.Dyn.Syst.2009}%
  \BibitemOpen
  \bibfield  {author} {\bibinfo {author} {\bibfnamefont {G.~A.}\ \bibnamefont {Gottwald}}\ and\ \bibinfo {author} {\bibfnamefont {I.}~\bibnamefont {Melbourne}},\ }\bibfield  {title} {\bibinfo {title} {On the {{Implementation}} of the 0--1 {{Test}} for {{Chaos}}},\ }\href {https://doi.org/10.1137/080718851} {\bibfield  {journal} {\bibinfo  {journal} {SIAM Journal on Applied Dynamical Systems}\ }\textbf {\bibinfo {volume} {8}},\ \bibinfo {pages} {129} (\bibinfo {year} {2009})}\BibitemShut {NoStop}%
\bibitem [{\citenamefont {Shortiss}\ \emph {et~al.}(2019)\citenamefont {Shortiss}, \citenamefont {Lingnau}, \citenamefont {Dubois}, \citenamefont {Kelleher},\ and\ \citenamefont {Peters}}]{ShortissOpt.Express2019}%
  \BibitemOpen
  \bibfield  {author} {\bibinfo {author} {\bibfnamefont {K.}~\bibnamefont {Shortiss}}, \bibinfo {author} {\bibfnamefont {B.}~\bibnamefont {Lingnau}}, \bibinfo {author} {\bibfnamefont {F.}~\bibnamefont {Dubois}}, \bibinfo {author} {\bibfnamefont {B.}~\bibnamefont {Kelleher}},\ and\ \bibinfo {author} {\bibfnamefont {F.~H.}\ \bibnamefont {Peters}},\ }\bibfield  {title} {\bibinfo {title} {Harmonic frequency locking and tuning of comb frequency spacing through optical injection},\ }\href {https://doi.org/10.1364/OE.27.036976} {\bibfield  {journal} {\bibinfo  {journal} {Optics Express}\ }\textbf {\bibinfo {volume} {27}},\ \bibinfo {pages} {36976} (\bibinfo {year} {2019})}\BibitemShut {NoStop}%
\bibitem [{\citenamefont {Wu}\ \emph {et~al.}(2025)\citenamefont {Wu}, \citenamefont {Peng}, \citenamefont {Yuan}, \citenamefont {Boscolo}, \citenamefont {Finot},\ and\ \citenamefont {Zeng}}]{Wu2025}%
  \BibitemOpen
  \bibfield  {author} {\bibinfo {author} {\bibfnamefont {X.}~\bibnamefont {Wu}}, \bibinfo {author} {\bibfnamefont {J.}~\bibnamefont {Peng}}, \bibinfo {author} {\bibfnamefont {B.}~\bibnamefont {Yuan}}, \bibinfo {author} {\bibfnamefont {S.}~\bibnamefont {Boscolo}}, \bibinfo {author} {\bibfnamefont {C.}~\bibnamefont {Finot}},\ and\ \bibinfo {author} {\bibfnamefont {H.}~\bibnamefont {Zeng}},\ }\bibfield  {title} {\bibinfo {title} {Unveiling the complexity of {{Arnold}}'s tongues in a breathing-soliton laser},\ }\href {https://doi.org/10.1126/sciadv.ads3660} {\bibfield  {journal} {\bibinfo  {journal} {Science Advances}\ }\textbf {\bibinfo {volume} {11}},\ \bibinfo {pages} {eads3660} (\bibinfo {year} {2025})}\BibitemShut {NoStop}%
\bibitem [{\citenamefont {Yuan}\ \emph {et~al.}(2018)\citenamefont {Yuan}, \citenamefont {Lin}, \citenamefont {Xiao},\ and\ \citenamefont {Fan}}]{YuanOptica2018}%
  \BibitemOpen
  \bibfield  {author} {\bibinfo {author} {\bibfnamefont {L.}~\bibnamefont {Yuan}}, \bibinfo {author} {\bibfnamefont {Q.}~\bibnamefont {Lin}}, \bibinfo {author} {\bibfnamefont {M.}~\bibnamefont {Xiao}},\ and\ \bibinfo {author} {\bibfnamefont {S.}~\bibnamefont {Fan}},\ }\bibfield  {title} {\bibinfo {title} {Synthetic dimension in photonics},\ }\href {https://doi.org/10.1364/OPTICA.5.001396} {\bibfield  {journal} {\bibinfo  {journal} {Optica}\ }\textbf {\bibinfo {volume} {5}},\ \bibinfo {pages} {1396} (\bibinfo {year} {2018})}\BibitemShut {NoStop}%
\bibitem [{\citenamefont {Bell}\ \emph {et~al.}(2017)\citenamefont {Bell}, \citenamefont {Wang}, \citenamefont {Solntsev}, \citenamefont {Neshev}, \citenamefont {Sukhorukov},\ and\ \citenamefont {Eggleton}}]{BellOptica2017}%
  \BibitemOpen
  \bibfield  {author} {\bibinfo {author} {\bibfnamefont {B.~A.}\ \bibnamefont {Bell}}, \bibinfo {author} {\bibfnamefont {K.}~\bibnamefont {Wang}}, \bibinfo {author} {\bibfnamefont {A.~S.}\ \bibnamefont {Solntsev}}, \bibinfo {author} {\bibfnamefont {D.~N.}\ \bibnamefont {Neshev}}, \bibinfo {author} {\bibfnamefont {A.~A.}\ \bibnamefont {Sukhorukov}},\ and\ \bibinfo {author} {\bibfnamefont {B.~J.}\ \bibnamefont {Eggleton}},\ }\bibfield  {title} {\bibinfo {title} {Spectral photonic lattices with complex long-range coupling},\ }\href {https://doi.org/10.1364/OPTICA.4.001433} {\bibfield  {journal} {\bibinfo  {journal} {Optica}\ }\textbf {\bibinfo {volume} {4}},\ \bibinfo {pages} {1433} (\bibinfo {year} {2017})}\BibitemShut {NoStop}%
\bibitem [{\citenamefont {Bal{\v c}ytis}\ \emph {et~al.}(2022)\citenamefont {Bal{\v c}ytis}, \citenamefont {Ozawa}, \citenamefont {Ota}, \citenamefont {Iwamoto}, \citenamefont {Maeda},\ and\ \citenamefont {Baba}}]{BalcytisSci.Adv.2022}%
  \BibitemOpen
  \bibfield  {author} {\bibinfo {author} {\bibfnamefont {A.}~\bibnamefont {Bal{\v c}ytis}}, \bibinfo {author} {\bibfnamefont {T.}~\bibnamefont {Ozawa}}, \bibinfo {author} {\bibfnamefont {Y.}~\bibnamefont {Ota}}, \bibinfo {author} {\bibfnamefont {S.}~\bibnamefont {Iwamoto}}, \bibinfo {author} {\bibfnamefont {J.}~\bibnamefont {Maeda}},\ and\ \bibinfo {author} {\bibfnamefont {T.}~\bibnamefont {Baba}},\ }\bibfield  {title} {\bibinfo {title} {Synthetic dimension band structures on a {{Si CMOS}} photonic platform},\ }\href {https://doi.org/10.1126/sciadv.abk0468} {\bibfield  {journal} {\bibinfo  {journal} {Science Advances}\ }\textbf {\bibinfo {volume} {8}},\ \bibinfo {pages} {eabk0468} (\bibinfo {year} {2022})}\BibitemShut {NoStop}%
\bibitem [{\citenamefont {Sridhar}\ \emph {et~al.}(2024)\citenamefont {Sridhar}, \citenamefont {Ghosh}, \citenamefont {Srinivasan}, \citenamefont {Miller},\ and\ \citenamefont {Dutt}}]{SridharNat.Phys.2024a}%
  \BibitemOpen
  \bibfield  {author} {\bibinfo {author} {\bibfnamefont {S.~K.}\ \bibnamefont {Sridhar}}, \bibinfo {author} {\bibfnamefont {S.}~\bibnamefont {Ghosh}}, \bibinfo {author} {\bibfnamefont {D.}~\bibnamefont {Srinivasan}}, \bibinfo {author} {\bibfnamefont {A.~R.}\ \bibnamefont {Miller}},\ and\ \bibinfo {author} {\bibfnamefont {A.}~\bibnamefont {Dutt}},\ }\bibfield  {title} {\bibinfo {title} {Quantized topological pumping in {{Floquet}} synthetic dimensions with a driven dissipative photonic molecule},\ }\href {https://doi.org/10.1038/s41567-024-02413-3} {\bibfield  {journal} {\bibinfo  {journal} {Nature Physics}\ }\textbf {\bibinfo {volume} {20}},\ \bibinfo {pages} {843} (\bibinfo {year} {2024})}\BibitemShut {NoStop}%
\bibitem [{\citenamefont {Lin}\ \emph {et~al.}(2016)\citenamefont {Lin}, \citenamefont {Xiao}, \citenamefont {Yuan},\ and\ \citenamefont {Fan}}]{LinNatCommun2016a}%
  \BibitemOpen
  \bibfield  {author} {\bibinfo {author} {\bibfnamefont {Q.}~\bibnamefont {Lin}}, \bibinfo {author} {\bibfnamefont {M.}~\bibnamefont {Xiao}}, \bibinfo {author} {\bibfnamefont {L.}~\bibnamefont {Yuan}},\ and\ \bibinfo {author} {\bibfnamefont {S.}~\bibnamefont {Fan}},\ }\bibfield  {title} {\bibinfo {title} {Photonic {{Weyl}} point in a two-dimensional resonator lattice with a synthetic frequency dimension},\ }\href {https://doi.org/10.1038/ncomms13731} {\bibfield  {journal} {\bibinfo  {journal} {Nature Communications}\ }\textbf {\bibinfo {volume} {7}},\ \bibinfo {pages} {13731} (\bibinfo {year} {2016})}\BibitemShut {NoStop}%
\bibitem [{\citenamefont {Lustig}\ \emph {et~al.}(2019)\citenamefont {Lustig}, \citenamefont {Weimann}, \citenamefont {Plotnik}, \citenamefont {Lumer}, \citenamefont {Bandres}, \citenamefont {Szameit},\ and\ \citenamefont {Segev}}]{LustigNature2019}%
  \BibitemOpen
  \bibfield  {author} {\bibinfo {author} {\bibfnamefont {E.}~\bibnamefont {Lustig}}, \bibinfo {author} {\bibfnamefont {S.}~\bibnamefont {Weimann}}, \bibinfo {author} {\bibfnamefont {Y.}~\bibnamefont {Plotnik}}, \bibinfo {author} {\bibfnamefont {Y.}~\bibnamefont {Lumer}}, \bibinfo {author} {\bibfnamefont {M.~A.}\ \bibnamefont {Bandres}}, \bibinfo {author} {\bibfnamefont {A.}~\bibnamefont {Szameit}},\ and\ \bibinfo {author} {\bibfnamefont {M.}~\bibnamefont {Segev}},\ }\bibfield  {title} {\bibinfo {title} {Photonic topological insulator in synthetic dimensions},\ }\href {https://doi.org/10.1038/s41586-019-0943-7} {\bibfield  {journal} {\bibinfo  {journal} {Nature}\ }\textbf {\bibinfo {volume} {567}},\ \bibinfo {pages} {356} (\bibinfo {year} {2019})}\BibitemShut {NoStop}%
\bibitem [{\citenamefont {Lustig}\ and\ \citenamefont {Segev}(2021)}]{LustigAdv.Opt.Photon.AOP2021}%
  \BibitemOpen
  \bibfield  {author} {\bibinfo {author} {\bibfnamefont {E.}~\bibnamefont {Lustig}}\ and\ \bibinfo {author} {\bibfnamefont {M.}~\bibnamefont {Segev}},\ }\bibfield  {title} {\bibinfo {title} {Topological photonics in synthetic dimensions},\ }\href {https://doi.org/10.1364/AOP.418074} {\bibfield  {journal} {\bibinfo  {journal} {Advances in Optics and Photonics}\ }\textbf {\bibinfo {volume} {13}},\ \bibinfo {pages} {426} (\bibinfo {year} {2021})}\BibitemShut {NoStop}%
\bibitem [{\citenamefont {Crameri}(2023)}]{Crameri2023}%
  \BibitemOpen
  \bibfield  {author} {\bibinfo {author} {\bibfnamefont {F.}~\bibnamefont {Crameri}},\ }\href {https://doi.org/10.5281/ZENODO.1243862} {\bibinfo {title} {Scientific colour maps}},\ \bibinfo {howpublished} {Zenodo} (\bibinfo {year} {2023})\BibitemShut {NoStop}%
\bibitem [{\citenamefont {Crameri}\ \emph {et~al.}(2020)\citenamefont {Crameri}, \citenamefont {Shephard},\ and\ \citenamefont {Heron}}]{CrameriNatCommun2020}%
  \BibitemOpen
  \bibfield  {author} {\bibinfo {author} {\bibfnamefont {F.}~\bibnamefont {Crameri}}, \bibinfo {author} {\bibfnamefont {G.~E.}\ \bibnamefont {Shephard}},\ and\ \bibinfo {author} {\bibfnamefont {P.~J.}\ \bibnamefont {Heron}},\ }\bibfield  {title} {\bibinfo {title} {The misuse of colour in science communication},\ }\href {https://doi.org/10.1038/s41467-020-19160-7} {\bibfield  {journal} {\bibinfo  {journal} {Nature Communications}\ }\textbf {\bibinfo {volume} {11}},\ \bibinfo {pages} {5444} (\bibinfo {year} {2020})}\BibitemShut {NoStop}%
\bibitem [{\citenamefont {Moille}\ \emph {et~al.}(2019{\natexlab{b}})\citenamefont {Moille}, \citenamefont {Li}, \citenamefont {Xiyuan},\ and\ \citenamefont {Srinivasan}}]{MoilleJ.RES.NATL.INST.STAN.2019}%
  \BibitemOpen
  \bibfield  {author} {\bibinfo {author} {\bibfnamefont {G.}~\bibnamefont {Moille}}, \bibinfo {author} {\bibfnamefont {Q.}~\bibnamefont {Li}}, \bibinfo {author} {\bibfnamefont {L.}~\bibnamefont {Xiyuan}},\ and\ \bibinfo {author} {\bibfnamefont {K.}~\bibnamefont {Srinivasan}},\ }\bibfield  {title} {\bibinfo {title} {{{pyLLE}}: {{A Fast}} and {{User Friendly Lugiato-Lefever Equation Solver}}},\ }\href {https://doi.org/10.6028/jres.124.012} {\bibfield  {journal} {\bibinfo  {journal} {Journal of Research of the NIST}\ }\textbf {\bibinfo {volume} {124}},\ \bibinfo {pages} {124012} (\bibinfo {year} {2019}{\natexlab{b}})}\BibitemShut {NoStop}%
\end{thebibliography}
%

\clearpage

\appendix   
\onecolumngrid
\renewcommand{\appendixpagename}{\Large\centering Supplementary Information: \mytitle}

\appendixpage
\setcounter{figure}{0}
\setcounter{equation}{0}
\renewcommand{\thesection}{S.\arabic{section}}
\renewcommand\thefigure{S.\arabic{figure}}    
\numberwithin{equation}{section}
\renewcommand\theequation{S.\arabic{equation}}

\section{Modified Lugiato Lefever Model}
\label{sup_sec:lle_model}

The Lugiato-Lefever equation models nonlinear dynamics in microring resonators~\cite{ChemboPhys.Rev.A2013}, and in the case of multiple pumping can be written as~\cite{TaheriEur.Phys.J.D2017}:
\begin{equation}
    \begin{split} 
    \label{sup_eq:mLLE}
    \frac{\partial a}{\partial t} &= 
        \left(-\frac{\kappa}{2} + i\delta \omega_{\mathrm{0}}\right)a 
            + i\sum_\mu D_{\mathrm{int}}(\mu)\tilde{a}(\mu,t)\mathrm{e}^{i\mu\theta} 
            - i\gamma  |a|^2 a + i\sqrt{\kappa_{\mathrm{ext}}P_{\mathrm{0}}}
        + i\sqrt{\kappa_\mathrm{ext}P_\mathrm{-}} \mathrm{exp}(i\varphi_\mathrm{ref} + i\mu_\mathrm{s}\theta)
    \end{split}
\end{equation}

\noindent with $\theta$ the azimuthal angle, $\tilde{a}(\mu,t)$ the Fourier transform of $a(\theta,t)$ with respect to $\theta$, and $D_\mathrm{int} = \omega_\mathrm{res}(\mu) - (\omega_0 + \omega_\mathrm{rep}^{(0)}\mu)$ the modified integrated dispersion (which in practice is determined from finite element method microring eigenmode simulations that match experimental data) with $\omega_\mathrm{res}(\mu)$ the resonant angular frequency of mode $\mu$ relative to the main pumped mode and $\omega_0$ the main pump angular frequency. %
$\omega_\mathrm{rep}^{(0)}$ is the unsynchronized repetition rate, and $\mu$ the relative mode number to the main pumped mode. For critical coupling, \qty{\kappa/2\pi =180}{\MHz} and $\kappa_\mathrm{ext} = \kappa/2$ define the total and coupling loss rates. The effective nonlinearity $\gamma =\nicefrac{2 \pi R n_2 \omega_0}{A_\mathrm{eff}(\omega_\mathrm{0}) c}$ uses \qty{R=23}{\um}, \qty{n_2=2.5\times 10^{-15}}{W\per\cm^2}, and \qty{A_\mathrm{eff}=0.464}{\um^2}. $P_0 = F_0^2/\kappa_\mathrm{ext}$, \qty{\omega_0/2\pi=283.5}{\THz}, and $\delta\omega_0$ represent the power, frequency and detuning of the main pump. The reference pump mode $\mu_\mathrm{s}$ has power $P_\mathrm{ref} = F_\mathrm{ref}^2/\kappa_\mathrm{ext}$ and offset $\partial\varphi_\mathrm{ref}/\partial t = \varpi= \omega_\mathrm{ref} - \omega_0 - \mu_\mathrm{s}\omega_\mathrm{rep}^{(0)} = \delta\omega_\mathrm{0} - \delta \omega_\mathrm{ref}+ D_\mathrm{int}(\mu_\mathrm{ref})$ from the closest soliton comb tooth.

The phase modulation can be directly implemented by adding a time-dependence to the phase term in Eq.~\ref{sup_eq:mLLE}: 
\begin{equation}
    \begin{split} 
    \label{sup_eq:ACmod_mLLE}
    \frac{\partial a}{\partial t} &= 
        \left(-\frac{\kappa}{2} + i\delta \omega_{\mathrm{0}}\right)a 
            + i\sum_\mu D_{\mathrm{int}}(\mu)\tilde{a}(\mu,t)\mathrm{e}^{i\mu\theta} 
            - i\gamma  |a|^2 a + i\sqrt{\kappa_{\mathrm{ext}}P_{\mathrm{0}}}
        + i\sqrt{\kappa_\mathrm{ext}P_\mathrm{-}} \mathrm{exp}\left[i\left(  \varphi_\mathrm{ref} + A\cos(\omega_\mathrm{rf}t) \right)+ i\mu_\mathrm{s}\theta \right]
    \end{split}
\end{equation}

In the small modulation amplitude regime, or when the reference pump detuning $\delta\omega_\mathrm{ref}$ from its closest comb tooth is less than the phase modulation frequency $\omega_\mathrm{rf}$, the Bessel expansion can be truncated to first order:

\begin{equation}
    \begin{split} 
    \label{sup_eq:ACmod_mLLE_expanded}
    \frac{\partial a}{\partial t} &= 
        \left(-\frac{\kappa}{2} + i\delta \omega_{\mathrm{0}}\right)a 
            + i\sum_\mu D_{\mathrm{int}}(\mu)\tilde{a}(\mu,t)\mathrm{e}^{i\mu\theta} 
            - i\gamma  |a|^2 a + i\sqrt{\kappa_{\mathrm{ext}}P_{\mathrm{0}}}
        + iJ_0(A)\sqrt{\kappa_\mathrm{ext}P_\mathrm{-}} \mathrm{exp}\left[i\varpi t + i\mu_\mathrm{s}\theta \right] \\
        &- J_1(A)\sqrt{\kappa_\mathrm{ext}P_\mathrm{-}} \mathrm{exp}\left[i\left( \varpi + \omega_\mathrm{rf}  \right)t+ i\mu_\mathrm{s}\theta \right] 
        - J_1(A)\sqrt{\kappa_\mathrm{ext}P_\mathrm{-}} \mathrm{exp}\left[i\left( \varpi - \omega_\mathrm{rf}  \right)t+ i\mu_\mathrm{s}\theta \right] +
    \end{split}
\end{equation}

\noindent which is equivalent to a DKS with three auxiliary lasers pumping at $\varpi$ and $\varpi \pm \omega_\mathrm{ref}$. At $\varpi \approx \pm \omega_\mathrm{ref}$, the phase of one auxiliary force aligns with the DKS, enabling direct synchronization at the carrier or sideband.

Beyond these cases, the system produces a multi-color soliton where auxiliary forces generate waves with locked group velocities to the DKS but different phase velocities. This results in distinct accumulated phases per round-trip and frequency components that share the comb tooth spacing $\omega_\mathrm{rep}$ but are offset by $\varpi$ and $\varpi\pm \omega_\mathrm{rf}$. These colors can be expanded into:

\begin{equation}
    \label{sup_eq:color_decomp}
    a = a_\mathrm{-}\mathrm{exp}\left[ i(\varpi-\omega_\mathrm{rf})t \right] +  a_\mathrm{0} + a_\mathrm{ref}\mathrm{exp}\left[ i\varpi t \right] + a_\mathrm{+} \mathrm{exp}\left[ i(\varpi+\omega_\mathrm{rf})t \right]
\end{equation}

As shown in~\cite{Menyuk2024,MoilleArXivParametric2024}, substituting~\cref{sup_eq:color_decomp} into~\cref{sup_eq:ACmod_mLLE_expanded} and retaining only phase matched terms yields the system of equations:
\begin{align}
    \frac{\partial a_-}{\partial t} =& 
        \left(- \frac{\kappa}{2} + i\varpi_-\right) a_- 
        + i\sum_\mu D_{\mathrm{int}}(\mu)\tilde{a}_-(\mu,t)\mathrm{e}^{i\mu\theta}
        - i\gamma \left(|a_{-}|^2 + 2|a_0|^2 + 2|a_{ref}|^2  + 2|a_{+}|^2\right)a_-  \nonumber\\
        &- i\gamma a_0^2a_\mathrm{ref}^*e^{-iW t}
        - J_1(A)\sqrt{\kappa_\mathrm{ext}P_\mathrm{ref}}\label{sup_eq:a-} \\
    \frac{\partial a_0}{\partial t} =& 
        \left(-\frac{\kappa}{2} + i\delta\omega_0\right) a_0 
        i\sum_\mu D_{\mathrm{int}}(\mu)\tilde{a}_0(\mu,t)\mathrm{e}^{i\mu\theta}
        - i\gamma \left(2|a_{-}|^2 + |a_0|^2 + 2|a_{ref}|^2  + 2|a_{+}|^2\right)a_0 \nonumber\\
        & - 2i\gamma a_0^*a_\mathrm{ref}a_{-}e^{iW t}
        + i\sqrt{\kappa_\mathrm{ext}P_0}\label{sup_eq:a0} \\
        \frac{\partial a_\mathrm{ref}}{\partial t} =& 
        \left(- \frac{\kappa}{2} + i\varpi_\mathrm{ref}\right) a_\mathrm{ref} 
        + i\sum_\mu D_{\mathrm{int}}(\mu)\tilde{a}_\mathrm{ref}(\mu,t)\mathrm{e}^{i\mu\theta}
        - i\gamma \left(2|a_{-}|^2 + 2|a_0|^2 + |a_{ref}|^2  + 2|a_{+}|^2\right)a_\mathrm{ref} \nonumber \\
        &- i\gamma a_\mathrm{0}^2a_{-}^*e^{-iW t}
        + iJ_0(A)\sqrt{\kappa_\mathrm{ext}P_\mathrm{ref}}\label{sup_eq:a+}\\
    \frac{\partial a_\mathrm{+}}{\partial t} =& 
        \left(- \frac{\kappa}{2} + i\varpi_\mathrm{+}\right) a_\mathrm{+} 
        + i\sum_\mu D_{\mathrm{int}}(\mu)\tilde{a}_+(\mu,t)\mathrm{e}^{i\mu\theta}
        - i\gamma \left(2|a_{-}|^2 + 2|a_0|^2 + 2|a_{ref}|^2  + |a_{+}|^2\right)a_\mathrm{+} \nonumber\\
        & - J_0(A)\sqrt{\kappa_\mathrm{ext}P_\mathrm{ref}}\label{sup_eq:aref}
\end{align}
with $W = 2\varpi - \omega_\mathrm{rf}$. The system synchronizes at $W=0$ (equivalently at $\varpi = \omega_\mathrm{rf}/2$) through parametric interaction between the DKS $a_0$, the reference $a_\text{ref}$, and its sideband $a_-$. Since $a_0$ and $a_-$ are driven at $\mu_s$ near a comb tooth, phase matching of the parametric interaction occurs at $\mu_s$, yielding the same OFD as direct or side-band KIS.

A similar principle governs the positive sideband and higher harmonics, demonstrating that sub-harmonic locking at $p\omega_\mathrm{rf}/m$ arises from parametric Kerr-induced synchronization.

\section{Adler equation model in static and phase modulated regimes}
\subsection{Adler equation in the static regime}
\label{sup_subsec:adler_dc}
In the unmodulated KIS, the phase offset between DKS and reference laser relates directly to their frequency difference:
\begin{equation}
    \label{sup_eq:ceo_offset}
    \begin{split}
        \frac{\partial \varphi}{\partial t} = \frac{\partial \varphi_\mathrm{ref} - \varphi_\mathrm{dks}}{\partial t} &= \varpi\\
                                             &= \omega_\mathrm{ref}  - \left( \mu_s \omega_\mathrm{rep}  +\omega_\mathrm{o} \right)
    \end{split}
\end{equation} 
With $\omega_\mathrm{ref}$ and $\omega_\mathrm{0}$ denoting the reference and main pump frequencies, $\mu_s$ representing the mode relative to the pumped mode at synchronization, and $\omega_\mathrm{rep} = v_g/R$ the repetition rate (where $v_g$ is the group velocity of the DKS and $R$ the microring radius), we normalize the phase to $\varphi_\mathrm{dks} = 0$. The optical frequency division (OFD) occurs at synchronization when $\partial\varphi/\partial t=0$
\begin{equation}
    \omega_\mathrm{rep} = \frac{\omega_\mathrm{ref} - \omega_\mathrm{0}}{\mu_\mathrm{s}}
\end{equation}

In the small amplitude approximation of the reference pump power $P_\mathrm{ref}$ outside synchronization, the DKS repetition rate remains unaffected by the reference pump~\cite{MoilleNature2023}, expressed as:

\begin{equation}
    \label{sup_eq:rep_rate}
    \frac{\partial\omega_\mathrm{rep}}{\partial t } = -\frac{1}{E_\mathrm{dks}} \sum_\mu D_1(\mu) \left(\frac{\partial \tilde{a}(\mu,t)}{\partial t}\tilde{a}^*(\mu,t) + \tilde{a}(\mu,t)\frac{\partial \tilde{a}^*(\mu,t)}{\partial t}  \right)
\end{equation}

Using the Fourier transform of the Lugiato Lefever equation (\cref{sup_eq:mLLE}, \cref{sup_eq:rep_rate}), we obtain:

\begin{equation}
    \label{sup_eq:dreprate}
    -\frac{1}{\kappa}\frac{\partial\omega_\mathrm{rep}}{\partial t } = -2D_1E_\mathrm{kis}\sin(\varphi) - \omega_\mathrm{rep} - D_1 \left( K_0 + K_\mathrm{NL} \right)
\end{equation}

\noindent where we introduce the normalized KIS energy $E_\mathrm{kis} =  \sqrt{E_0(\mu_s) P_\mathrm{ref}\kappa_\mathrm{ext}}/\kappa E_\mathrm{dks}$, with $E_\mathrm{dks}= \int_{-\pi}^{\pi}|a|^2 \mathrm{d}\theta$ as the total soliton energy and its order of magnitude is \qty{\approx10^{-9}}{J}, and $E_0(\mu_s) = |\tilde{a}(\mu_s)|^2$ which order of magnitude is \qty{\approx10^{-14}}{J} for a dispersive wave. For the weakly dispersive resonator, $D_1(\mu=0)\approx D_1(\mu = \mu_s) = D_1$. The pump frequency shift and the self-phase modulation frequency shift are: 
\begin{align}
    K_0 &= 2\sqrt{\kappa_\mathrm{ext}}\frac{P_0}{\kappa E_\mathrm{dks}}\\
    K_\mathrm{NL} &= \frac{\gamma}{\kappa E_\mathrm{dks} D_1 }\sum_\mu D_1(\mu)\left(\tilde{a}^*\left( \mu ,t \right) \sum_{\alpha,\beta}\tilde{a}(\alpha, t)\tilde{a}^*(\beta)\tilde{a}(\alpha-\beta+\mu, t)  - c.c \right)
\end{align}

\noindent From the steady state of \cref{sup_eq:dreprate}, one find that $D_1(K_0 + K_\mathrm{NL}) \approx \omega_\mathrm{rep}$ since $E_\mathrm{kis}\ll 1$

\cref{sup_eq:ceo_offset} relates the phase offset to the temporal variation in repetition rate at fixed optical frequencies:
\begin{equation}
    \label{sup_eq:d2phidt2}
    \frac{\partial\omega_\mathrm{rep}}{\partial t} = -\frac{1}{\mu_s}\frac{\partial^2\varphi}{\partial t^2}
\end{equation}

Using \cref{sup_eq:ceo_offset,sup_eq:dreprate,sup_eq:d2phidt2}, and noting that $\omega_\mathrm{ref} - \omega_\mathrm{0} - D_1\mu_s(K_0+K_\mathrm{NL})\approx \delta\omega_\mathrm{ref} + \mu_s\omega_\mathrm{rep} - \mu_s\omega_\mathrm{rep}$ is the frequency offset $\delta\omega_\mathrm{ref}$ of the reference relative to its nearest comb tooth at $\mu_s$, we obtain the normalized Adler equation:
\begin{equation}
    \label{sup_eq:dc_adler}
    \beta\frac{\partial\varphi^2}{\partial\tau^2} + \frac{\partial\varphi}{\partial\tau} = \alpha  + \sin(\varphi)
\end{equation}

The normalization in~\cref{tab:sup_param} establishes the Adler model's universality, extending beyond Kerr-induced synchronization to driven damped pendula~\cite{CoulletAmericanJournalofPhysics2005} and Josephson junctions~\cite{ValizadehJNMP2008}.

\cref{sup_eq:dc_adler} can be rewritten as a system of first order differential equations, yielding equation (1) of the main text where $U = \alpha\varphi + \cos(\varphi) + 1$.
\begin{equation}
     \label{sup_eq:dc_adler_system}
     \begin{dcases}
         \dfrac{\partial\varphi}{\partial \tau} = \varOmega\\
         \beta\frac{\partial \varOmega}{\partial \tau} = - \varOmega +  \frac{\partial U}{\partial \varphi}    
     \end{dcases}
 \end{equation}

\begin{table}[H]
    \captionsetup{width=1\linewidth}
    \caption{\label{tab:sup_param}\textbf{Physical parameters used in the normalized Adler equation --} for KIS, the parameters are extracted from the LLE and the experiment.}
    \renewcommand{\arraystretch}{2.5}
    \begin{center}
        \begin{tabularx}{1\linewidth}{c >{\centering}X >{\centering}X >{\centering}X}    
            \toprule
              & Damped pendulum under constant torque~\cite{CoulletAmericanJournalofPhysics2005} & Josephson junction~\cite{ValizadehJNMP2008} & Kerr-induced synchronization~\cite{MoilleNature2023}\\
            \midrule
            $\displaystyle \tau = \varOmega_0 t$  %
                & %
                $\displaystyle \frac{mgl}{v}$
                & %
                $\displaystyle \frac{2e}{\hbar}I_\mathrm{c} Rt$
                & $\displaystyle  2|\mu_s|  D_1 E_\mathrm{kis} t$ \\
            $\displaystyle \Gamma$  %
                & %
                $\displaystyle \frac{mgl}{I}$
                & %
                $\displaystyle \frac{1}{RC}$
                & $\kappa$ \\
            $\displaystyle \beta = \frac{\varOmega_0}{\Gamma}$  %
                &  %
                $\displaystyle \frac{I}{v} $
                & %
                $\displaystyle  \frac{2 e}{\hbar}I_cR^2 C$
                & $\displaystyle  \frac{2|\mu_s|  D_1 E_\mathrm{kis}}{\kappa} $ \\
            $\displaystyle\alpha = \frac{\varOmega_\mathrm{ext}}{\varOmega_0}$  %
            &  %
            $\displaystyle \frac{v}{mgl}\varOmega_m$ 
            &  %
            $\displaystyle \frac{I_\mathrm{ext}}{I_\mathrm{c}}$
            & %
            $\displaystyle \frac{\delta\omega_\mathrm{ref}}{2\mu_s  D_1 E_\mathrm{kis}}$\\
            \bottomrule
        \end{tabularx}
    \end{center}
    \renewcommand{\arraystretch}{1}
\end{table}

\subsection{Phase modulated Adler equation}
\label{sup_subsec:adler_ac}
The phase modulation of the reference, introduced in~\cref{sup_eq:ACmod_mLLE},changes $\varphi_\mathrm{ref} \rightarrow \varphi_\mathrm{ref} + A\cos\left( \omega_\mathrm{rf}t \right)$. The derivation proceeds as above, incorporating this reference phase modulation:

\begin{equation}
    \label{sup_eq:ac_dphidt}
    \begin{split}
        \frac{\partial \varphi}{\partial t} = \frac{\partial \varphi_\mathrm{ref} - \varphi_\mathrm{dks}}{\partial t} &= \varpi - \omega_\mathrm{rf}A\mathrm{sin}\left( \omega_\mathrm{rf}t \right)\\
                            &= \omega_\mathrm{ref} -  \omega_\mathrm{rf}A\mathrm{sin}\left( \omega_\mathrm{rf}t \right) - \left( \mu_s \omega_\mathrm{rep}  +\omega_\mathrm{o} \right)
    \end{split}
\end{equation}
which can be recast as: 
\begin{equation}
    \omega_\mathrm{rep} = \frac{1}{\mu_s}\left[ \omega_\mathrm{ref} -\omega_0 - \omega_\mathrm{rf}A\mathrm{sin}\left( \omega_\mathrm{rf}t \right) -\frac{\partial\varphi}{\partial t} \right]
\end{equation}
from which we obtain:
\begin{equation}
    \label{sup_eq:ac_d2phidt2}
    \frac{1}{\kappa}\frac{\partial\omega_\mathrm{rep}}{\partial t} = - \frac{1}{\kappa\mu_s}\frac{\partial^2\varphi}{\partial t^2} - \frac{A}{\kappa\mu_s}\omega_\mathrm{rf}^2\cos\left( \omega_\mathrm{rf} t\right)
\end{equation}

Using \cref{sup_eq:dreprate,sup_eq:ac_dphidt,sup_eq:ac_d2phidt2}, we derive the phase-modulated Adler equation:
\begin{equation}
    \label{sup_eq:a_adler}
    \beta\frac{\partial\varphi^2}{\partial\tau^2} + \frac{\partial\varphi}{\partial\tau} = \alpha  + \sin(\varphi + A\cos\left( \varOmega_\mathrm{ext} \tau \right)) 
        + A \eta \cos(\varOmega_\mathrm{ext} + \delta\phi)
\end{equation}

\noindent with $\eta = \omega_\mathrm{rf}\sqrt{\kappa^2 + \omega_\mathrm{rf}^2}/\mu_s\kappa\varOmega_0$ and $\delta\phi = \tan^{-1}\left( \omega_\mathrm{rf} /\kappa \right)$. Under conditions where the synchronization mode is far from the main pump $\mu_s \gg 1$, the term $\eta \approx -0.042$ becomes negligible for working amplitude depth $A$ near unity, given \qty{\omega_\mathrm{rf}/2\pi=200}{\MHz}, \qty{\varOmega_\mathrm{0}/2\pi=120}{\MHz}, \qty{\kappa/2\pi=100}{\MHz}, and $\mu=-88$. The equation can be expanded using the Bessel expansion for small detuning up to the first sideband:

\begin{equation}
    \label{sup_eq:a_adlerBessel}
    \beta\frac{\partial\varphi^2}{\partial\tau^2} + \frac{\partial\varphi}{\partial\tau} = \alpha  + J_0(A)\sin(\varphi) \pm J_1(A)\sin\left( \varOmega_\mathrm{ext}t \right)
\end{equation}

We can express this as a system of first-order differential equations, as shown in equation (2) of the main text:

\begin{equation}
    \label{sup_eq:ac_adler}
    \begin{dcases}
        \frac{\partial\varphi}{\partial \tau} = \varOmega \\
        \frac{\partial\psi}{\partial \tau}  = \varOmega_\mathrm{ext}\\
        \beta\frac{\partial \varOmega}{\partial \tau}  = - \varOmega +  \alpha + J_0(A)\sin(\varphi) \pm J_1(A)\sin(\psi)\\
    \end{dcases}
\end{equation}

\section{Experimental Setup}
\label{sup_sec:exp_setup}
\begin{figure}[h]
    \centering
    \includegraphics{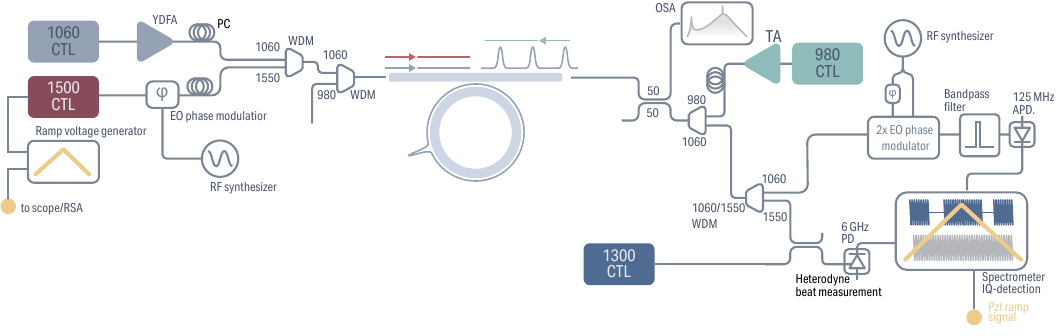}
    \caption{\label{supfig:setup} \textbf{Experimental Setup}
    A \qty{283}{\THz} continuously tunable laser (1060 CTL) is amplified by a ytterbium-doped fiber amplifier (YDFA) and aligned to the transverse electric (TE) mode with a polarization controller (PC) to pump the microring resonator. A \qty{306}{\THz} laser (980 CTL), counterpropagating in transverse magnetic (TM) polarization, thermally stabilizes the microring and enables adiabatic access to the DKS. The \qty{193}{\THz} reference laser (1500 CTL) passes through a lithium-niobate electro-optic (EO) phase modulator driven by a radiofrequency (RF) synthesizer. An optical spectrum analyzer (OSA) measures the comb spectrum. Two cascaded EO phase modulators translate adjacent comb teeth into a detectable beat to measure the DKS repetition rate. The CEO offset beat is obtained by splitting the comb at \qty{1550}{\nm} with a wavelength demultiplexer (WDM) and detecting the interference with a fast photodiode.
    }
\end{figure}

\indent Our experimental resonator consists of a \qty{670}{\nm} thick \ce{Si3N4} microring resonator in \ce{SiO2} with a \qty{23}{\um} radius and \qty{850}{\nm} ring width.
A \qty{460}{\nm} wide bus waveguide wrapped in a \qty{17}{\um} pulley and offset by a \qty{600}{\nm} gap provides for dispersion-less coupling~\cite{MoilleOpt.Lett.2019}. %
The experimental setup [\cref{supfig:setup}] employs a \qty{1060}{\nm} continuously tunable laser (CTL), amplified by an ytterbium-doped fiber amplifier (YDFA). %
When pumped at \qty{282.3}{\THz} with \qty{150}{\mW} on-chip power, a dissipative Kerr soliton (DKS) state emerges in the fundamental transverse electric (TE) resonator. %
A counterpropagating, cross-polarized counter-propagative \qty{307.6}{\THz} laser amplified using an tapered amplifier (TA) provides thermal stabilization for adiabatic DKS access\cite{ZhangOptica2019, ZhouLightSciAppl2019}, while minimizing nonlinear pump mixing. %
We employ wavelength demultiplexers (WDMs) at both ports to prevent cross-injection between the cooler and main pump. %
The \qty{1550}{nm} CTL reference laser, modulated by a lithium-niobate phase modulator driven by a radiofrequency synthesizer, enters the resonator in TE polarization via WDM combination with the main pump. %
A 50~\% tap at the output directs light to an OSA for microcomb spectrum analysis. %
A WDM at the other 50~\% output separates wavelengths near \qty{1100}{nm}, which undergo electro-optical (EO) modulation at \qty{\varOmega_\mathrm{RF}/2\pi=17.84420}{\GHz}. %
The EO-generated sidebands of adjacent DKS comb teeth fill the DKS repetition rate over $N=56$ EOcomb teeth, measuring {\qty{\omega_\mathrm{rep} = N \times \varOmega_\mathrm{RF} + \delta\omega_\mathrm{beat} = 2\pi\times(999.275241 + \delta\omega_\mathrm{beat})}{\GHz}} (the plus sign is configuration dependent and verified experimentally), where $\delta\omega_\mathrm{beat}$ is the EOcomb beat note detected by a \qty{50}{\MHz} avalanche photodiode (APD), as in~\cite{StonePhys.Rev.Lett.2020,MoilleNature2023}. 

From this measurement, we extract the detuning between the reference laser and the comb tooth, assuming a calibrated reference detuning against the unsynchronized comb $\varpi$. %
Since the repetition rate is disciplined at the OFD rate $\mu_s$, we have $\Delta\omega_\mathrm{rep} = \Delta\delta_\mathrm{beat} = \varpi \mu_s^{-1}$. %
From such relation, it is obvious that the comb tooth is captured by the reference as no beat can be recorded between them two since $\partial (\mu_s \Delta\omega_\mathrm{rep} - \varpi)/\partial\varpi = 0$. %
The system presents a low dispersion, thus the group velocity of the different colors outside of KIS are bound through cross-phase modulation~\cite{WangOptica2017a,MoilleNat.Photon.2025}. %
Therefore the previous relationship extends beyond the KIS window such that $\Delta\omega_\mathrm{ceo} = \partial (\mu_s \Delta\omega_\mathrm{rep} - \varpi)/\partial\varpi \equiv \partial \varOmega/\partial\alpha$. %
Hence, from the recording of the instantaneous repetition rate variation, we are able to recreate the mapping of a fixed point presented in Figure~2.

The remaining WDM output measures the homodyne beat between the reference and its nearest comb tooth. %
Both signals undergo time-domain processing using an electrical spectral analyzer triggered to the reference frequency sweep voltage ramp, converting the photodiode's IQ temporal signal to spectrograms and measuring instantaneous frequency. %
The repetition rate is bounded by the APD bandwidth, hence its spectrogram enables to retrieve the low-passed filter repetition rate disciplining. %
However, the photodiode used for the homodyne is fast enough to resolve the modulation frequency, hence enables for the latter enables for the reconstruction of the spectrogram shown in Figure~3.

\section{Randomness results}

We performed randomness tests according to the NIST SP 800-22 standard~\cite{Rukhin2010}, which includes a suite of statistical tests to evaluate the randomness of binary sequences and provided through a python package~\cite{RandomGithub}. %
The tests were applied to the DKS repetition rate time series, which was sampled at \qty{280}{\MHz}  (\textit{i.e.} as close as possible to the photon lifetime with the instrument in use) for \qty{400}{\ms} and yielded a total of $112\times10^6$ samples. %
The data are processed such that they are converted to a binary sequence based on which attractor (carrier or side-band) the repetition rate is pinned to. %
The results are summarized in \cref{suptab:monobit} and \cref{suptab:chi2}, confirming the randomness of the chaotic repetition rate hopping induced by the phase-modulation of the Kerr-induced synchronization.

\begin{table}[htbp]
    \centering
    \caption{\label{suptab:monobit} Frequency monobit test of the repetition rate hopping time series}
    \begin{tabular}{c C{4cm}c}
            \toprule
            \textbf{Test} & \textbf{P-Value} & \textbf{Conclusion} \\
            \midrule
            Frequency (Monobit) Test & 0.6142348491858782 & Random \\
            \bottomrule
    \end{tabular}
    
\end{table}

\begin{table}[htpb]
    \caption{\label{suptab:chi2} Chi-squared test of the repetition rate hopping time series}
    \begin{tabular}{c C{4cm} C{4cm}c}
            \toprule
            \textbf{State} & \textbf{Chi-Squared} & \textbf{P-Value} & \textbf{Conclusion} \\      
            \midrule
            - 4 & 9.357578281776533 & 0.0956238214109253 & Random \\
            - 3 & 12.49469090909091 & 0.0286034169822541 & Random \\
            - 2 & 4.856341189674524 & 0.4336630733098357 & Random \\
            - 1 & 2.272727272727273 & 0.8102640548687199 & Random \\
            +1 & 1.0 & 0.9625657732472964 & Random \\
            +2 & 3.3434343434343434 & 0.6471987509656730 & Random \\
            +3 & 6.476363636363637 & 0.2625845685276736 & Random \\
            +4 & 11.664192949907235 & 0.0396903797798626 & Random \\
            \bottomrule
        \end{tabular}
        
\end{table}

 \end{document}